\documentclass[a4paper, twocolumn,
superscriptaddress,
preprintnumbers,
nofootinbib, nobibnotes,
amsmath, amssymb,
aps, prd,
fleqn, floatfix, 10pt]{revtex4-2}
\usepackage[english]{babel}
\usepackage{amsmath,graphicx,xcolor}
\definecolor{xlinkcolor}{cmyk}{1,1,0,0}
\usepackage[bookmarks=true, pdfnewwindow=true, colorlinks=true, linkcolor=xlinkcolor, citecolor=xlinkcolor, filecolor=xlinkcolor, urlcolor=xlinkcolor, final=true]{hyperref}

\newcommand{\dd}{{\rm d}}
\newcommand{\xs}{\sigma}
\newcommand{\CR}{\text{CR}}
\newcommand{\thr}{\text{thr}}
\newcommand{\muphNucl}{\mu(\leftarrow \gamma A)}
\newcommand{\leading}{\text{L}}

\begin{document}

\preprint{INR-TH-2025-022}

\title[l]{Readdressing the contribution of photonuclear reactions\texorpdfstring{\\}{} to the muon content of extensive air showers: a heuristic approach}

\author{Nickolay S. Martynenko}
\thanks{Contact author: martynenko@inr.ac.ru}
\affiliation{Faculty of Physics, Lomonosov Moscow State University,\\1-2 Leninskie Gory, Moscow 119991, Russia}
\affiliation{Institute for Nuclear Research of the Russian Academy of Sciences,\\
60th October Anniversary Prospect 7a, Moscow 117312, Russia}

\date{Submitted to \textit{Physical Review D} on December 16, 2025. Published on April 17, 2026}

\begin{abstract}
The indirect ground-based observations of cosmic rays through extensive air showers in modern experiments typically involve the use of Monte Carlo simulations to determine the characteristics of the primary particles. These simulations necessitate assumptions about particle interactions at energies that have not yet been experimentally probed, which introduces systematic uncertainties in key observables, particularly the number of muons. Current research on this uncertainty primarily focuses on hadronic interaction models, the dominant source of muon production. This study presents an approach that takes into account another significant mechanism for muon generation: photonuclear reactions. A robust heuristic technique has been developed to estimate the contribution of these interactions to the total number of muons over a wide range of extensive air shower parameters (including primary particle type, energy, and slant atmospheric depth) and photonuclear interaction models, with an absolute percentage error on the order of \(10\%\) in the estimated number of muons. Furthermore, several potential applications of the suggested method in relation to modern challenges in extensive air shower physics are discussed.

\end{abstract}
\maketitle

\section{Introduction}
\label{sec:intro}
Measuring the mass composition of cosmic rays is crucial for understanding their acceleration and propagation mechanisms. However, for cosmic rays with energies above \(E_{\CR} \sim 10^{15}\,\text{eV}\), only indirect ground-based measurements via extensive air showers (EAS) are feasible, as the flux of such high-energy cosmic rays is too low for direct detection. State-of-the-art measurements typically require sophisticated reconstruction methods, which utilize Monte Carlo (MC) simulations of both the EAS development and the detector response~\cite{Kampert:2012}. These simulations, in turn, require assumptions about particle interactions in energy ranges beyond those probed by collider experiments.

One of the key EAS observables related to the mass composition of cosmic rays is the number of muons. Notably, several independent experiments reveal an intriguing disagreement between their data and MC simulations. The observed number of muons significantly exceeds the MC expectation for the same energy of the primary particle, and the discrepancy increases with energy. This problem is called the “muon puzzle”. For a more detailed discussion, see~\cite{Albrecht:2022}, which reviews recent experimental and theoretical advances on the muon puzzle, and~\cite{Velazquez:2023}, which summarizes key experimental results in a meta-analysis.

Research on the muon puzzle typically focuses on tuning models of high-energy hadronic interactions to reconcile collider data, EAS observations, and corresponding MC simulations~\cite{Albrecht:2022}. However, despite numerous attempts, a definitive theoretical explanation of the muon puzzle is still lacking.

This study does not aim to develop a suitable model for hadronic interactions. Instead, it examines the relationship between the number of muons and the photonuclear reaction (PNR) cross section at high energies, focusing on the role of PNRs in transferring energy from the electromagnetic to the hadronic cascade. In these reactions, energetic photons (with energy \(E_{\gamma A} \gtrsim 10^{8}\,\text{eV}\) in the nucleus rest frame) are absorbed by atmospheric nuclei, triggering the production of secondary hadrons. The subsequent decay or further interactions of these secondaries ultimately lead to muon production, thereby enhancing the total number of muons in EAS.

On the one hand, it is believed that the contribution of PNRs to the number of muons is subdominant compared to hadronic interactions, with the corresponding muon fraction ranging from \(\sim 2\%\) to \(\sim11\%\), depending on the cosmic ray mass composition and the muon detection threshold~\cite{Muller:2018}. On the other hand, the high-energy PNR cross section is not well constrained by laboratory measurements. As reported by the Particle Data Group \cite{PDG:2024}, the PNR cross section has not yet been measured for incident photon laboratory-frame energies above \(E_{\gamma A} \sim 10^{13}\,\text{eV}\) for proton targets (\(\sqrt{s} \sim 10^{11}\,\text{eV}\)). This lack of data necessitates model extrapolations, leading to significant theoretical ambiguities. To date, a wide range of experimentally allowed theoretical models have been proposed for high-energy PNRs, as discussed in Chapter~6 of the review by Pancheri and Srivastava~\cite{Pancheri:2016}, each predicting distinct asymptotic behaviors.

Furthermore, there is an additional systematic uncertainty arising from the transformation of PNR cross sections from proton targets to atmospheric nuclei (such as nitrogen \(^{14}{\rm N}\) and oxygen \(^{16}{\rm O}\)). Recent studies have proposed non-trivial data-driven transformations \cite{Kossov:2002,Morejon:2019}, while EAS MC simulation frameworks, such as CORSIKA \cite{Heck:1998} and CONEX \cite{Pierog:2004,Bergmann:2006}, implement simple power-law transformations.

Previous studies have considered the uncertainty in the PNR cross section in relation to the EAS muon content. For instance, \cite{Ochs:1985,Drees:1988,Wdowczyk:1990,Fletcher:1991,Krys:1991,Dumora:1992} discussed the possibility of a large and rapid increase in the high-energy PNR cross section, motivated by observations of an anomalous muon rate from Cygnus X-3 in the mid-1980s. Despite the assumption of rather exotic cross-section behavior, these studies demonstrated that analytical methods can effectively link variations in the PNR cross section to changes in the muon number in photon-induced EAS. Subsequent studies \cite{Risse:2005,Cornet:2015} investigated the effect of the PNR cross-section’s asymptotic behavior on key observables of photon-induced EAS, particularly the muon content, which is critical for distinguishing them from nucleus-induced events by comparing the results of corresponding MC simulations. These results remain relevant for distinguishing between photon-induced and nucleus-induced EAS events. This is particularly important for the search for ultra-high-energy photons, as photon-induced EAS are expected to have a lower muon content compared to nucleus-induced events, making the number of muons a key discriminant, as discussed in the review \cite{PierreAuger:2022}.

However, previous studies have generally been limited in scope, focusing on specific instances of PNR cross section models and photon-induced EAS. This work addresses these limitations by proposing a novel approach that combines the advantages of previous efforts and extends them further. The goal is to develop a heuristic model that describes the change in the EAS muon content due to variations in the assumed PNR cross section. Unlike previous analytical approaches, the proposed model does not require a specific form of the PNR cross section. Instead, it assumes that high-energy PNRs in EAS occur infrequently enough that their contribution to the number of muons can be expressed as a linear functional of the PNR cross section. The model introduces heuristic functions that parameterize the energy and slant-depth dependence of muon production seen in MC simulations of relevant EAS processes. Both photon- and nucleus-initiated EAS are considered within a unified framework, ensuring a consistent description across the full spectrum of primary particle types. Based on these principles, this study presents a practical and widely applicable tool for studying systematic uncertainties in the number of muons arising from uncertainties in the PNR cross section at high energies.

The rest of the paper is organized as follows. Table~\ref{tab:notation} summarizes the main notation used in this manuscript.
\begin{table*}[htb]
    \caption{Notation used in this article (including placeholder symbols \(\square\), \(\triangle\)). All energies are in the laboratory frame. For photon-induced EAS, the convention of \(E_{\CR} \mapsto E_{\gamma}\) and \(A_{\CR} \mapsto 0\) is used.}
    \label{tab:notation}
    \vspace{\baselineskip}
    
    \begin{tabular}{lll}
        \textbf{Notation} & \textbf{Valid Range} & \textbf{Definition} \\
        \hline
        \(\square\) & \((-\infty, +\infty)\) & Arbitrary scalar variable \\
        \(\triangle\) & \((-\infty, +\infty)\) & Scalar variable (not to be confused with \(\Delta\)) \\
        \(\mathcal{I}\{\square\}\) & \((-\infty, +\infty)\) & Intercept parameter of the linear model for \(\square\) \\
        \(\mathcal{S}\{\square, \triangle\}\) & \((-\infty, +\infty)\) & Slope parameter of the linear model for \(\square\) \\
            & & with respect to variable \(\triangle\) (i.e., \(\dd \square / \dd \triangle\)) \\
        \(E_{\CR}\) & \([10^{15}, 10^{19}]\,\text{eV}\) & Primary cosmic-ray nucleus energy \\
        \(A_{\CR}\) & \(\{1, 4, \dots, 56\}\) & Primary cosmic-ray nucleus mass number \\
        \(N_{\gamma}\) & \((0, +\infty)\) & Average number of photons from hadron-air interactions or hadronic decays \\
        \(E_{\gamma}\) & \((0, E_{\CR}]\) & Energy of a photon (hadronic origin in nucleus-induced EAS; \\
             & & primary in photon-induced EAS)\\
        \(X_{\gamma}\) & \([0, 2000]\,\text{g cm}^{-2}\) & Slant depth of a photon \\
        \(E_{\gamma A}\) & \([E_{\gamma A}^{\thr}, E_{\gamma}]\) & Energy of an incident photon inducing a high-energy PNR \\
        \(E_{\gamma A}^{\thr}\) & \([10^{11}, 10^{19}]\,\text{eV}\) & Threshold energy for “high-energy” PNR (\(E_{\gamma A} \geq E_{\gamma A}^{\thr}\)) \\
        \(A\) & \(\{14,16, 40\}\) & PNR target nucleus mass number (\(^{14}{\rm N}\), \(^{16}{\rm O}\), \(^{40}{\rm Ar}\)) \\
        \(\xs_{\gamma p}\) & \((0, +\infty)\,\text{mb}\) & PNR cross section for proton target \\
        \(\xs_{\gamma A}\) & \((0, +\infty)\,\text{mb}\) & PNR cross section for nucleus of mass number \(A\) \\
        \(\bar{\xs}_{\gamma A}\) & \((0, +\infty)\,\text{mb}\) & PNR cross section averaged over atmospheric nuclei \\
        \(N_{\mu}\) & \((0, +\infty)\) & Average number of muons at a fixed slant atmospheric depth \(X_{\mu}\) \\
        \(N_{\muphNucl}\) & \((0, +\infty)\) & Number of muons from PNRs with \(E_{\gamma A} \geq E_{\gamma A}^{\thr}\) \\
        \(E_{\mu}\) & \([E^{\thr}_{\mu}, E_{\CR})\) & Energy of a muon \\
        \(E^{\thr}_{\mu}\) & \(=10^9\,\text{eV}\) & Minimum muon energy required to contribute to \(N_{\mu}\) \\
        \(X_{\mu}\) & \([0, 2000]\,\text{g cm}^{-2}\) & Slant depth of a muon \\
        \hline
    \end{tabular}
\end{table*}
Section~\ref{sec:method} develops the general approach for constructing the linear functional and explicitly states the key assumptions. Section~\ref{sec:monte-carlo} describes the MC simulations used to fit the linear functional and validate its applicability. Section~\ref{sec:analytical} proposes a heuristic parametrization of the linear functional and optimizes the parameters based on the MC simulation results. Section~\ref{sec:disc} provides a quantitative evaluation of the best-fitting model performance, including comparisons with simulation data and a discussion of its limitations. Finally, Section~\ref{sec:concl} summarizes the main results of this work.

\section{Method}
\label{sec:method}
For simplicity, only the longitudinal dimension (parametrized by slant depth) is considered. Perfectly efficient muon detection with a fixed lower energy threshold of \(E_\mu \geq E_{\mu}^{\thr}=10^9\,\text{eV}\) is assumed, which is typical for EAS experiments \cite{Muller:2018}. Only the average number of muons is considered, without accounting for the underlying distribution or fluctuations.

Hence, in this analysis, the average muon content of any nucleus-induced EAS is characterized by three parameters: the primary particle energy \(E_{\CR}\), its mass number \(A_{\CR}\), and the slant depth of muon detection \(X_\mu\).

It is assumed that the average number of muons produced by high-energy PNRs within EAS (hereafter denoted PN\(\mu\)) is a linear functional of the PNR cross section:
\begin{equation}
    \label{eq:linear-functional}
    \begin{aligned}
    &N_{\muphNucl}(E_{\gamma A}^{\thr}, E_{\CR}, A_{\CR}, X_{\mu})=\\
    &\int\limits_{E_{\gamma A}^{\thr}}^{E_{\CR}}\dd E_{\gamma A}\,\frac{\delta N_{\muphNucl}}{\delta \bar{\xs}_{\gamma A}(E_{\gamma A})} \times \bar{\xs}_{\gamma A}(E_{\gamma A}),
    \end{aligned}
\end{equation}
where the integral accounts for all PNRs above the threshold energy \(E_{\gamma A}^{\thr}\). The threshold is set at \(E_{\gamma A}^{\thr} \geq 10^{10}\,\text{eV}\), excluding the resonant region (\(E_{\gamma A} < 10^{10}\,\text{eV}\)) from the scope of this study. The PNR cross section for proton targets in this energy range is tightly constrained by experimental data, leaving little room for model variation~\cite{PDG:2024}. However, the uncertainty in the PNR cross section for atmospheric nuclei is non‑negligible and constitutes a significant contribution to the overall systematic uncertainty. A detailed analysis of this contribution is deferred here for future work. The goal is to find a simple heuristic approximation for the Green's function \(\delta N_{\muphNucl}/\delta \bar{\xs}_{\gamma A}(E_{\gamma A})\). From Eq.~\eqref{eq:linear-functional}, it follows that this function can be expressed as:
\begin{equation}
    \begin{aligned}
    \frac{\delta N_{\muphNucl}}{\delta \bar{\xs}_{\gamma A}(E_{\gamma A})}&=
    [\bar{\xs}_{\gamma A}(E_{\gamma A})]^{-1} \times \\
    &\times E_{\gamma A}^{-1}\frac{\dd N_{\muphNucl}(E_{\gamma A}, E_{\CR}, A_{\CR}, X_{\mu})}{\dd \ln E_{\gamma A}}.
    \end{aligned}
\end{equation}

The Superposition model proposed by~\cite{Gorjunov:1962} is adopted for PN\(\mu\), i.e., an EAS of energy \(E_{\CR}\) induced by a nucleus of mass number \(A_{\CR}\) is identified with \(A_{\CR}\) identical proton-induced EAS of energy \(A^{-1}_{\CR}E_{\CR}\) (in terms of their PN\(\mu\) content):
\begin{equation}
    \begin{aligned}
    &N_{\muphNucl}(E_{\gamma A}^{\thr}, E_{\CR}, A_{\CR}, X_{\mu}) =\\
    &A_{\CR}\times N_{\muphNucl}\left(E_{\gamma A}^{\thr}, A^{-1}_{\CR}E_{\CR}, 1, X_{\mu}\right).
    \end{aligned}
\end{equation}

The problem thus reduces to finding an analytical approximation for the function \(G(E_{\gamma A}, E_{\CR}, X_{\mu})\) defined as follows:
\begin{equation}
    G=[\bar{\xs}_{\gamma A}(E_{\gamma A})]^{-1}\frac{\dd N_{\muphNucl}(E_{\gamma A}, E_{\CR}, 1, X_{\mu})}{\dd \ln E_{\gamma A}}.
\end{equation}
By construction, \(G(E_{\gamma A}, E_{\CR}, X_{\mu})\) is independent of the assumed PNR cross section model, as the proportionality of \(dN_{\muphNucl}\) to \(\bar{\xs}_{\gamma A}(E_{\gamma A})\) cancels out the cross section dependence.

To parametrize \(G(E_{\gamma A}, E_{\CR}, X_{\mu})\), the EAS development is conditionally divided into two stages: (i) hadronic interactions (e.g., neutral pion production and decay) produce high-energy photons (hereafter denoted as hadronic photons); (ii) these hadronic photons either directly induce PNRs or produce electromagnetic cascades, within which PNRs may occur. These reactions partially transfer photon energy back to the hadronic cascade, ultimately contributing to PN\(\mu\) content of the EAS.

Stage (i) depends on the hadronic cascade development at high energies and high altitudes, while stage (ii) is primarily determined by the electromagnetic cascade development and the assumed PNR cross section model. Based on this observation, the following factorization of \(G(E_{\gamma A}, E_{\CR}, X_{\mu})\) is adopted:
\begin{equation}
    \label{eq:G-factorization}
    \begin{aligned}
    G&=\int\limits_{E_{\gamma A}}^{E_{\CR}}\frac{\dd E_{\gamma}}{E_{\gamma}}\int\limits_{0}^{X_{\mu}}\dd X_{\gamma}\Bigg[\frac{\dd N_{\gamma}(E_{\gamma}, X_{\gamma}, E_{\CR})}{\dd\ln E_{\gamma}\,\dd X_{\gamma}}\times\\ &\times \Gamma(E_{\gamma A}, E_{\gamma},X_{\mu}-X_{\gamma})\Bigg].
    \end{aligned}
\end{equation}
The first factor in the integrand describes the distribution of hadronic photons over their energies \(E_{\gamma}\) and slant depths of their parent reactions \(X_{\gamma}\), and the second factor \(\Gamma(E_{\gamma A}, E_{\gamma}, X_{\mu}-X_{\gamma})\) is the counterpart of \(G(E_{\gamma A}, E_{\CR}, X_{\mu})\) for photon-induced EAS, with \(X_{\mu} - X_{\gamma}\) representing the remaining slant depth available for cascade development after the photon is produced at \(X_{\gamma}\):
\begin{equation}
    \Gamma = [\bar{\xs}_{\gamma A}(E_{\gamma A})]^{-1}\frac{\dd N_{\muphNucl}(E_{\gamma A}, E_{\gamma}, 0, X_{\mu}-X_{\gamma})}{\dd \ln E_{\gamma A}}.
\end{equation}

The leading contribution to the number of PN\(\mu\) is expected to arise from the most energetic hadronic photons in the EAS, as their higher energy increases the expected number of high-energy PNRs and subsequent PN\(\mu\) production. Therefore, only the first few most energetic interactions in the EAS are considered to estimate the hadronic photon distribution. Section~\ref{sec:monte-carlo} provides a detailed discussion of this approximation.

\section{Monte Carlo Simulations}
\label{sec:monte-carlo}
To probe the hadronic and electromagnetic cascade contributions to the function \(G(E_{\gamma A}, E_{\CR}, X_{\mu})\), this study utilizes CONEX version 7.80~\cite{Pierog:2004,Bergmann:2006}, incorporating the EPOS LHC-R model for high-energy hadronic interactions, UrQMD 1.3 for low-energy hadronic interactions, and EGS4 for electromagnetic interactions. The atmospheric composition assumed in EGS4 includes three atomic species: nitrogen \(^{14}{\rm N}\), oxygen \(^{16}{\rm O}\), and argon \(^{40}{\rm Ar}\). The code uses a simple power-law scaling of \(\sigma_{\gamma A}/\sigma_{\gamma p}=A^{0.91}\) for the PNR cross section~\cite{Heck:1998}.

CONEX combines MC simulation of high-energy interactions with a fast numerical solution of cascade equations (CE). In this study, the default transition threshold \(E_{\text{MC\(\rightarrow\)CE}}/E_{\CR}=5\times10^{-3}\) is chosen to balance computational efficiency with accuracy in modeling the cascade development.

Instead of assuming a specific cosmic-ray spectrum or varying the zenith angle \(\theta\), vertical EAS (\(\theta = 0\)) with fixed primary energy are simulated. The extended longitudinal profiles are recorded from the slant depth of \(X=0\) to \(X=2000\,\text{g cm}^{-2}\) with a step of \(10\,\text{g cm}^{-2}\), sufficient to resolve the longitudinal development of muons in EAS. This enables the study of inclined EAS under the assumption that cascade development is universal with respect to slant depth, as supported by previous studies~\cite{Andringa:2012,Cazon:2022}.

\subsection{High-energy hadronic cascade development}
\label{subsec:monte-carlo:hadronic}
To estimate \(\dd N_{\gamma}/(\dd \ln E_{\gamma}\,\dd X_{\gamma})\), the “Leading interactions tree” option in CONEX is utilized. This method tracks the most energetic secondary particles at each interaction step, as these dominate the production of high-energy hadronic photons.

When enabled, the option records a data set including the atmospheric slant depth, the secondary particle content, and the energies of the secondaries for the first interaction in an EAS. Then, an analogous data set is recorded for the subsequent interaction of the leading particle (i.e., the most energetic particle among the secondaries) with the atmosphere, continuing iteratively.

In addition, photon-air interactions in EGS4 are manually disabled to isolate hadronic photon production. This is justified because electromagnetic cascade photons are already accounted for by \(\Gamma\). Also, since neutral pions decay almost immediately (lifetime \(\sim 8.4\times10^{-17}\)\,s~\cite{PDG:2024}), any recorded \(\pi^0\) of energy \(E_{\pi^0}\) at depth \(X_{\pi^0}\) is approximated as producing two photons of energy \(E_{\gamma}= E_{\pi^0}/2\) at the same depth \(X_{\gamma} = X_{\pi^0}\). This approximation holds in the ultra-relativistic limit relevant for EAS. These photon pairs are then added to the initial photon content recorded in the Leading interactions tree.

The EPOS LHC-R model implies that the multiplicity \(\mathcal{M}_{p+\text{air}}\) of the proton interactions with air grows rapidly with center-of-mass energy and reaches \(\mathcal{M}_{p+\text{air}} \sim 700\) at the highest energies, while the average energy fraction of the leading particle decreases steeply with energy from \(\varepsilon_{p+\text{air}} \sim 0.5\) to \(\varepsilon_{p+\text{air}} \sim 0.3\), as shown in~\cite{Pierog:2025}.

Thus, the lower bound for the average energy \(E_{\not \leading,\,1}\) of a non-leading particle produced in the first interaction can be estimated as:
\begin{equation}
    \operatorname{l.b.}[E_{\not \leading,\,1}] \sim \min\limits_{E_{\CR}}\left[\frac{1-\varepsilon_{p+\text{air}}}{\mathcal{M}_{p+\text{air}}} E_{\CR}\right] \sim  10^{-3} \times E_{\CR}.
\end{equation}
The upper bound for the leading particle energy \(E_{ \leading,\,k}\) after \(k\) interactions is:
\begin{equation}
    \operatorname{u.b.}[E_{\leading,\,k}] \sim \max\limits_{E_{\CR}}\left[(\varepsilon_{p+\text{air}})^k E_{\CR}\right] \sim 0.5^k \times E_{\CR}.
\end{equation}
After several interactions, \(\operatorname{u.b.}[E_{\leading,\,k}] \propto 0.5^k\) approaches \(\operatorname{l.b.}[\bar{E}_{\not\leading,\,1}]\). Interactions of the non-leading particles are not recorded in our approach, and therefore, to avoid exceeding of accuracy, only the first \(k_{\max}=\operatorname{round}(\log_{0.5}10^{-3})=10\) leading interactions are recorded.

It should be noted that the remaining part of the hadronic cascade, while not carrying the most energetic interactions, might make a significant contribution to the function \(G(E_{\gamma A},\,E_{\CR},\,X_{\mu})\). This work actually estimates only the contribution from the Leading interactions tree, namely \(\dd N_{\gamma\leading}/(\dd \ln E_{\gamma}\,\dd X_{\gamma})\) and \(G_{\leading}(E_{\gamma A},\,E_{\CR},\,X_{\mu})\). To account for this remaining contribution, which is not directly simulated here, an empirical correction factor \(\kappa_G>1\) is introduced, defined by
\begin{equation}
    \label{eq:fudge-factor}
    G(E_{\gamma A},\,E_{\CR},\,X_{\mu}) = \kappa_G \times G_{\leading}(E_{\gamma A},\,E_{\CR},\,X_{\mu}).
\end{equation}
A ballpark value of \(\kappa_G = 5\) is set for all subsequent calculations, based on preliminary comparisons with full cascade simulations. However, this value should be used with caution because the correction factor is sensitive to the model of high-energy hadronic interactions. For a detailed discussion, see Appendix~\ref{appendix:fudge-factor}.

Simulations cover the primary proton energy range \(10^{15}\,\text{eV}\leq E_{\CR}\leq10^{19}\,\text{eV}\) with a step of \(0.5\,\text{dex}\) (hereafter, dex is a contraction of “decimal exponent”: \(1\,\text{dex}\) corresponds to a factor of \(10\)). For each primary energy, 8192 distinct EAS are modeled to ensure statistical stability. These are randomly split into two equal groups (4096 each) to form the “training” and “test” datasets. The “training” group is used to fit the target distribution \(\dd N_{\gamma\leading}/(\dd\ln E_{\gamma}\,\dd X_{\gamma})\). The “test” group is then used to estimate the quality of the fit.

\subsection{Electromagnetic cascade development}
\label{subsec:monte-carlo:electromagnetic}
To estimate \(\Gamma(E_{\gamma A}, E_{\gamma}, X_{\mu})\), photon-induced EAS MC simulations are used. The laboratory energy range for PNR incident photons, \(10^{10}\,\text{eV}\leq E_{\gamma A}\leq 10^{19}\,\text{eV}\), is divided into bins with a width of \(0.2~\text{dex}\). In each simulation, a specific bin is selected. Using an algorithm similar to that described in Appendix B of ref.~\cite{Martynenko:2024}, the PNR cross section (and, by-product, the total cross section encoded by the photon mean free path) is modified in the EGS4 tables so that:
\begin{equation}
    \bar{\xs}_{\gamma A}(E_{\gamma A})=
    \begin{cases}
    \bar{\xs}_{\gamma A}(E_{\gamma A}\mid \text{EGS4}),\quad&\text{\(E_{\gamma A} \in\) bin;}\\
    0,\quad&\text{otherwise.}
    \end{cases}
\end{equation}
This modification ensures that PNRs occur only for photons in the selected bin, as the cross section is artificially suppressed to zero outside it. Thus, all PN\(\mu\) in an EAS are produced by PNRs within the given bin.

However, these are not all the muons. Sub-leading muon production mechanisms within the electromagnetic cascade, such as direct muon pair production, also contribute. To subtract this sub-leading contribution, additional MC simulations are performed with the PNRs disabled, i.e.,
\begin{equation}
    \bar{\xs}_{\gamma A}(E_{\gamma A}) = 0.
\end{equation}

Simulations cover the primary photon energy range \(10^{10}\,\text{eV}\leq E_{\gamma} \leq 10^{19}\,\text{eV}\) with a step of \(0.5~\text{dex}\). For each primary photon energy and for each bin, \(1024\) distinct EAS are modeled to ensure statistical stability. These are randomly split into two equal groups (\(512\) each) to form the “training” and “test” datasets. The “training” group is used to fit the target function \(\Gamma(E_{\gamma A}, E_{\gamma}, X_{\mu})\). The  “test” group is then used to estimate the quality of the fit.

In addition, to justify the linearity assumption, an extra bin of \(0 < E_{\gamma A} \leq 10^{10}\,\text{eV}\) is included in simulations. It is observed that the muon longitudinal profiles summed over the \(E_{\gamma A}\) bins (with sub-leading processes contribution subtracted except for the extra bin) is consistent with those for the standard photon-induced EAS of the same primary photon energy, as expected from the linearity assumption. This agreement holds except at high altitudes (several tens of \(\text{g~cm}^{-2}\) in slant atmospheric depth), where the average number of PNRs per one EAS is too low, leading to large statistical fluctuations in the average number of produced muons (relative uncertainty \(\gtrsim 50\%\)).

\subsection{Extensive air showers with modified photonuclear cross section}
\label{subsec:monte-carlo:modified-cs}
To evaluate our model for the PN\(\mu\) number prediction task, an additional set of EAS is modeled. An auxiliary smoothing function is defined:
\begin{equation}
    \operatorname{smooth}(\square) =
    \frac{1 + \tanh\left(10\log_{10}\square - 5\right)}{2},
\end{equation}
where the coefficients 10 and 5 set the transition steepness and midpoint, respectively, and a parametric family of piecewise continuous modified PNR cross sections is introduced:
\begin{equation}
    \label{eq:parametricXsection}
    \begin{aligned}
    &\bar{\xs}_{\gamma A}(E_{\gamma A}\mid E_{\gamma A}^{\thr}, a) = \bar{\xs}_{\gamma A}(E_{\gamma A}\mid \text{EGS4}) \times \\
    &\times \begin{cases}
        1 + (a-1) \times \operatorname{smooth}\left(\frac{E_{\gamma A}}{E_{\gamma A}^{\thr}}\right), & E_{\gamma A} \geq E_{\gamma A}^{\thr}, \\
        1, & E_{\gamma A} < E_{\gamma A}^{\thr},
    \end{cases}
    \end{aligned}
\end{equation}
where \(a \geq 0\) is the modified-to-EGS4 PNR cross section ratio at high energies.

To cover the entire range of relevant primary particle masses and energies, the standard EGS4 PNR cross section is used, and a set of EAS is modeled for five primary particles: photons \(\gamma\), protons \(^{1}\mathrm{H}\), helium nuclei \(^{4}\mathrm{He}\), oxygen nuclei \(^{16}\mathrm{O}\), and iron nuclei \(^{56}\mathrm{Fe}\). In the simulations, the primary particle energy range \(10^{15}\,\text{eV} \leq E_{\gamma},\,E_{\CR} \leq 10^{19}\,\text{eV}\) is scanned with a step of \(0.5~\text{dex}\). For each primary particle type and energy, \(1024\) distinct EAS are modeled to ensure statistical stability. To avoid data leakage, the random seeds used for these EAS simulations differ from those used in the training and testing datasets discussed previously.

Next, the PNR cross section in the EGS4 tables is modified according to Eq.~\eqref{eq:parametricXsection}, using \(E_{\gamma A}^{\thr} = 10^{10}\,\text{eV}\) and \(a=2\), and the same set of EAS simulations is repeated. The difference in the resulting muon content is expected to be entirely due to PN\(\mu\), enabling its prediction via Eq.~\eqref{eq:linear-functional}.

Additionally, the cross section amplitude parameter range \(a \in \{1/8,\, 1/4,\, 1/2,\, 2,\, 4,\, 8\}\) spanning diverse variations from strong suppression (\(a = 1/8\)) to strong enhancement (\(a = 8\)) is scanned at a fixed threshold \(E_{\gamma A}^{\thr} = 10^{10}\,\text{eV}\), as well as the threshold range \(10^{10}\,\text{eV} \leq E_{\gamma A}^{\thr} \leq 10^{16}\,\text{eV}\) with a \(2.0~\text{dex}\) step at fixed \(a=2\). For each PNR cross section model and each primary particle energy (within the same range), \(1024\) photon-induced and \(1024\) proton-induced EAS simulations are performed.

\section{Parametrization and Fitting}
\label{sec:analytical}
This section introduces a simple analytical parametrization for both \(\dd N_{\gamma\leading}/(\dd \ln E_{\gamma}\,\dd X_{\gamma})\) and \(\Gamma(E_{\gamma A}, E_{\gamma}, X_{\mu})\). The parametrization forms are chosen to reflect the physical behavior of the distributions while maintaining computational feasibility.

Next, the parameters are optimized to fit the training datasets (see Section~\ref{sec:monte-carlo} for simulation details). The quality of the fit is subsequently evaluated using the test datasets, ensuring robustness against overfitting.

Then, Eq.~\eqref{eq:G-factorization} is employed to numerically compute the desired function \(G_{\leading}(E_{\gamma A}, E_{\CR}, X_{\mu})\) on a predefined grid. 

Finally, the resulting integral is parametrized using an appropriate functional form. The parameters of this final parametrization are optimized to match the numerically evaluated function values.

\subsection{Hadronic photons from the leading interactions}
\label{subsec:analytical:hadron-leading-tree}

The following parametrization is adopted for the distribution of high-energy photons produced by hadronic interactions in the Leading interactions tree:
\begin{equation}
    \label{eq:dNgamma/dlnEgammadXgamma-parametrization}
    \begin{aligned}
    \frac{\dd N_{\gamma\leading}(E_{\gamma}, X_{\gamma}, E_{\CR})}{\dd\ln E_{\gamma}\,\dd X_{\gamma}} &= \Theta(E_{\CR}-E_{\gamma})\times\Theta(X_{\gamma}) \times \\
    &\times \frac{\mu_{[N_{\gamma\leading}]}}{\varsigma_{[\ln E_{\gamma}]}\,\mu_{[X_{\gamma}]}} \times \frac{\dd P}{\dd\epsilon_{\gamma}\,\dd x_{\gamma}},
    \end{aligned}
\end{equation}
where \(\Theta\) is the Heaviside step function, \(\mu_{[\square]}\) and \(\varsigma_{[\square]}\) represent, respectively, the expectation and the standard deviation of the quantity “\(\square\)” indicated in brackets for the EAS induced by a proton of energy \(E_{\CR}\), \(\dd P/(\dd \epsilon_{\gamma}\,\dd x_{\gamma})\) is the parametric two-dimensional probability density function describing the distribution of normalized quantities
\begin{equation}
  \epsilon_{\gamma} = \frac{\ln E_{\gamma} - \mu_{[\ln E_{\gamma}]}}{\varsigma_{[\ln E_{\gamma}]}}, \quad
  x_{\gamma} = \frac{X_{\gamma}}{\mu_{[X_{\gamma}]}}.
\end{equation}
Note that by construction, these normalized variables satisfy \(\mu_{[\epsilon_{\gamma}]} = 0\), \(\varsigma_{[\epsilon_{\gamma}]} = 1\), and \(\mu_{[x_{\gamma}]} = 1\).

We assume that \(\ln \mu_{[N_{\gamma\leading}]}\), \(\mu_{[\ln E_{\gamma}]}\), \(\mu_{[X_{\gamma}]}\), and \(\varsigma_{[\ln E_{\gamma}]}\) depend linearly on \(\ln E_{\CR}\). A linear regression model is used to fit the corresponding intercept and slope parameters to the training data, with uncertainties estimated via standard error propagation.

For the probability density, we adopt the following data-driven parametrization:
\begin{equation}
    \label{eq:dP-depsilon-dx-parametrization}    
    \frac{\dd P}{\dd\epsilon_{\gamma}\,\dd x_{\gamma}} = \frac{\exp(-x_{\gamma})}{\tilde\varsigma_{[\epsilon_{\gamma}\mid x_{\gamma}]}} \times f\left(\frac{\epsilon_{\gamma}-\tilde\mu_{[\epsilon_{\gamma}\mid x_{\gamma}]}}{\tilde\varsigma_{[\epsilon_{\gamma}\mid x_{\gamma}]}}\right),
\end{equation}
where \(\tilde\mu_{[\epsilon_{\gamma}\mid x_{\gamma}]}\) and \(\tilde\varsigma_{[\epsilon_{\gamma}\mid x_{\gamma}]}\) are, respectively, the conditional expectation and standard deviation of \(\epsilon_{\gamma}\) at fixed \(x_{\gamma}\). By definition,
\begin{equation}
    f(\square) = 
    \begin{cases}
        \frac{3}{4\sqrt{5}}\left(1-\frac{1}{5}\square^2\right), & |\square| \leq \sqrt{5}, \\
        0, & |\square| > \sqrt{5},
    \end{cases}
\end{equation}
is a quadratic probability density function with expectation \(0\) and standard deviation \(1\). This form captures the essential shape of the empirical distribution, while being numerically straightforward.

The adopted parametrization guarantees \(\mu_{[x_{\gamma}]} = 1\) by construction. Non-trivial constraints follow from the conditions \(\mu_{[\epsilon_{\gamma}]} = 0\) and \(\varsigma_{[\epsilon_{\gamma}]} = 1\), which lead to the integral equations. The first condition yields:
\begin{equation}
    \label{eq:conditional-expectation-constraint}
    \int\limits_0^{\infty} \dd x_{\gamma}\, e^{-x_{\gamma}} \tilde\mu_{[\epsilon_{\gamma}\mid x_{\gamma}]} = 0,
\end{equation}
while the second condition results in:
\begin{equation}
    \label{eq:conditional-dispersion-constraint}
    \int\limits_0^{\infty} \dd x_{\gamma}\, e^{-x_{\gamma}} \left( [\tilde\mu_{[\epsilon_{\gamma}\mid x_{\gamma}]}]^2 + [\tilde\varsigma_{[\epsilon_{\gamma}\mid x_{\gamma}]}]^2 \right) = 1.
\end{equation}

It is assumed that \(\tilde\mu_{[\epsilon_{\gamma}\mid x_{\gamma}]}\) is a linear function of \(x_{\gamma}\), and \(\tilde\varsigma_{[\epsilon_{\gamma}\mid x_{\gamma}]}\) is a linear function of \(\tilde\mu_{[\epsilon_{\gamma}\mid x_{\gamma}]}\). Under these assumptions, Eq.~\eqref{eq:conditional-expectation-constraint} implies the relation:
\begin{equation}
    \mathcal{S}\{\tilde\mu_{[\epsilon_{\gamma}\mid x_{\gamma}]}, x_{\gamma}\} = -\mathcal{I}\{\tilde\mu_{[\epsilon_{\gamma}\mid x_{\gamma}]}\},
\end{equation}
where the slope \(\mathcal{S}\) and the intercept \(\mathcal{I}\) are defined in Table~\ref{tab:notation}.
Combining Eqs.~\eqref{eq:conditional-expectation-constraint} and \eqref{eq:conditional-dispersion-constraint} leads to the expression:
\begin{equation}
    \begin{aligned}
    &\mathcal{I}\{\tilde\mu_{[\epsilon_{\gamma}\mid x_{\gamma}]}\}^2 \times \left(1 + \mathcal{S}\{\tilde\varsigma_{[\epsilon_{\gamma}\mid x_{\gamma}]}, \tilde\mu_{[\epsilon_{\gamma}\mid x_{\gamma}]}\}^2\right) + \\
    &+ \mathcal{I}\{\tilde\varsigma_{[\epsilon_{\gamma}\mid x_{\gamma}]}\}^2 = 1,
    \end{aligned}
\end{equation}
from which the intercept is derived as:
\begin{equation}
    \label{eq:intercept-conditional-expectation}
    \mathcal{I}\{\tilde\mu_{[\epsilon_{\gamma}\mid x_{\gamma}]}\} = \sqrt{\frac{1 - \mathcal{I}\{\tilde\varsigma_{[\epsilon_{\gamma}\mid x_{\gamma}]}\}^2}{1 + \mathcal{S}\{\tilde\varsigma_{[\epsilon_{\gamma}\mid x_{\gamma}]}, \tilde\mu_{[\epsilon_{\gamma}\mid x_{\gamma}]}\}^2}}.
\end{equation}

The resulting model contains two free parameters for the probability density: \(\mathcal{I}\{\tilde\varsigma_{[\epsilon_{\gamma}\mid x_{\gamma}]}\}\) and \(\mathcal{S}\{\tilde\varsigma_{[\epsilon_{\gamma}\mid x_{\gamma}]}, \tilde\mu_{[\epsilon_{\gamma}\mid x_{\gamma}]}\}\). These parameters are determined by maximizing the likelihood function derived from Eq.~\eqref{eq:dP-depsilon-dx-parametrization} using the \((\epsilon_{\gamma}, x_{\gamma})\) training dataset.

Analysis shows that the best-fitting value of \(\mathcal{I}\{\tilde\varsigma_{[\epsilon_{\gamma}\mid x_{\gamma}]}\}\) approaches \(1.0\), which maps \(\mathcal{I}\{\tilde\mu_{[\epsilon_{\gamma}\mid x_{\gamma}]}\}\) to \(0.0\) according to Eq.~\eqref{eq:intercept-conditional-expectation}. To ensure numerical stability in the fitting procedure, the parameter \(-\ln(1 - \mathcal{I}\{\tilde\varsigma_{[\epsilon_{\gamma}\mid x_{\gamma}]}\})\) is optimized instead of \(\mathcal{I}\{\tilde\varsigma_{[\epsilon_{\gamma}\mid x_{\gamma}]}\}\).

The optimization process minimizes the negative log-likelihood function normalized to the training set size. The implementation uses \texttt{scipy.optimize.minimize} with the COBYQA algorithm~\cite{Ragonneau:2022,Ragonneau:2025}. Parameter uncertainties are estimated through analytical derivation of the Hessian matrix for the normalized negative log-likelihood function based on the probability density in Eq.~\eqref{eq:dP-depsilon-dx-parametrization}. The uncertainty for each parameter is obtained as the square root of the corresponding diagonal element of the inverse Hessian matrix at the best-fitting point.

\begin{figure*}[htb]
    \centering
    \includegraphics[height=0.46\linewidth]{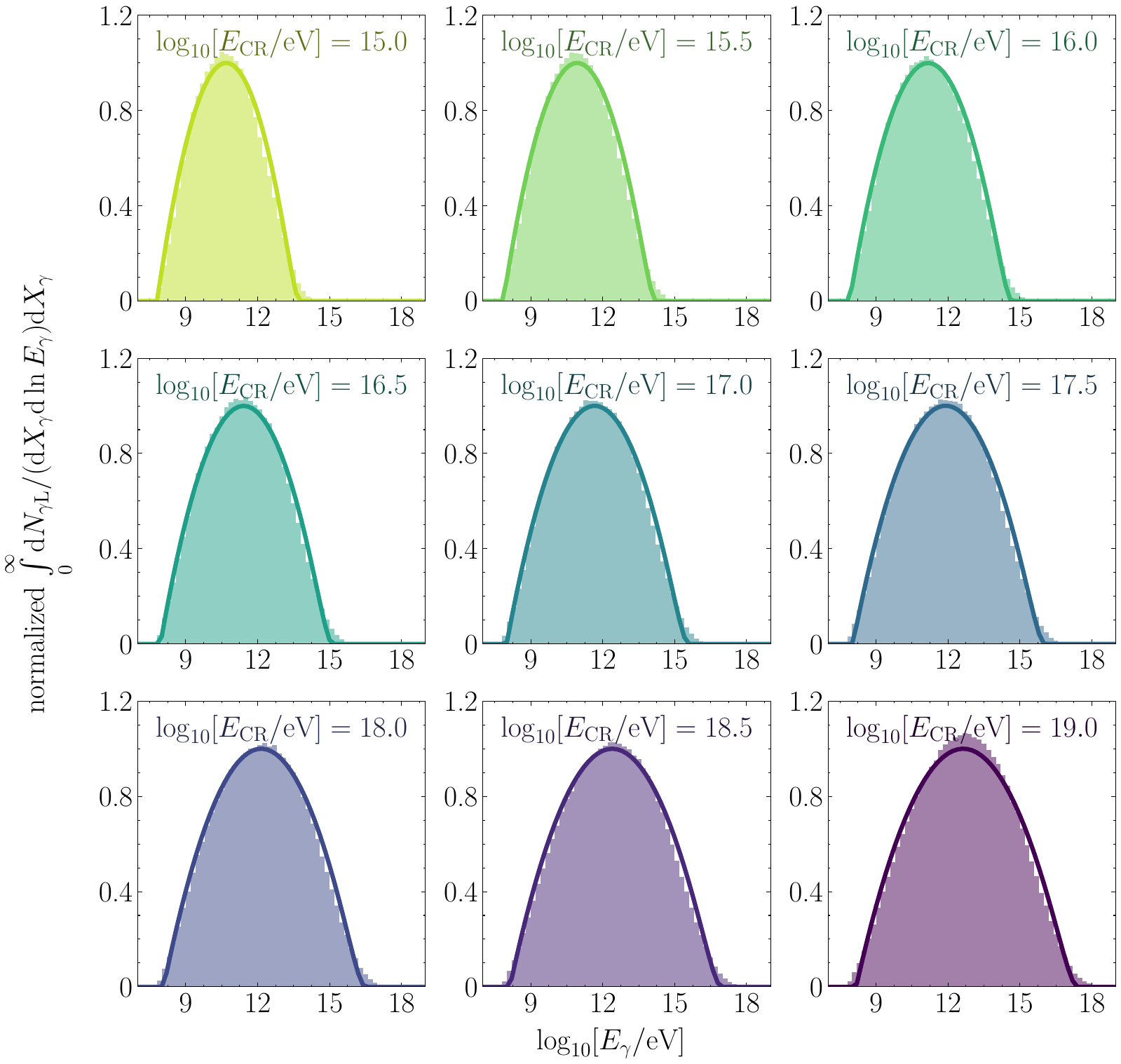}
    \hfill
    \includegraphics[height=0.46\linewidth]{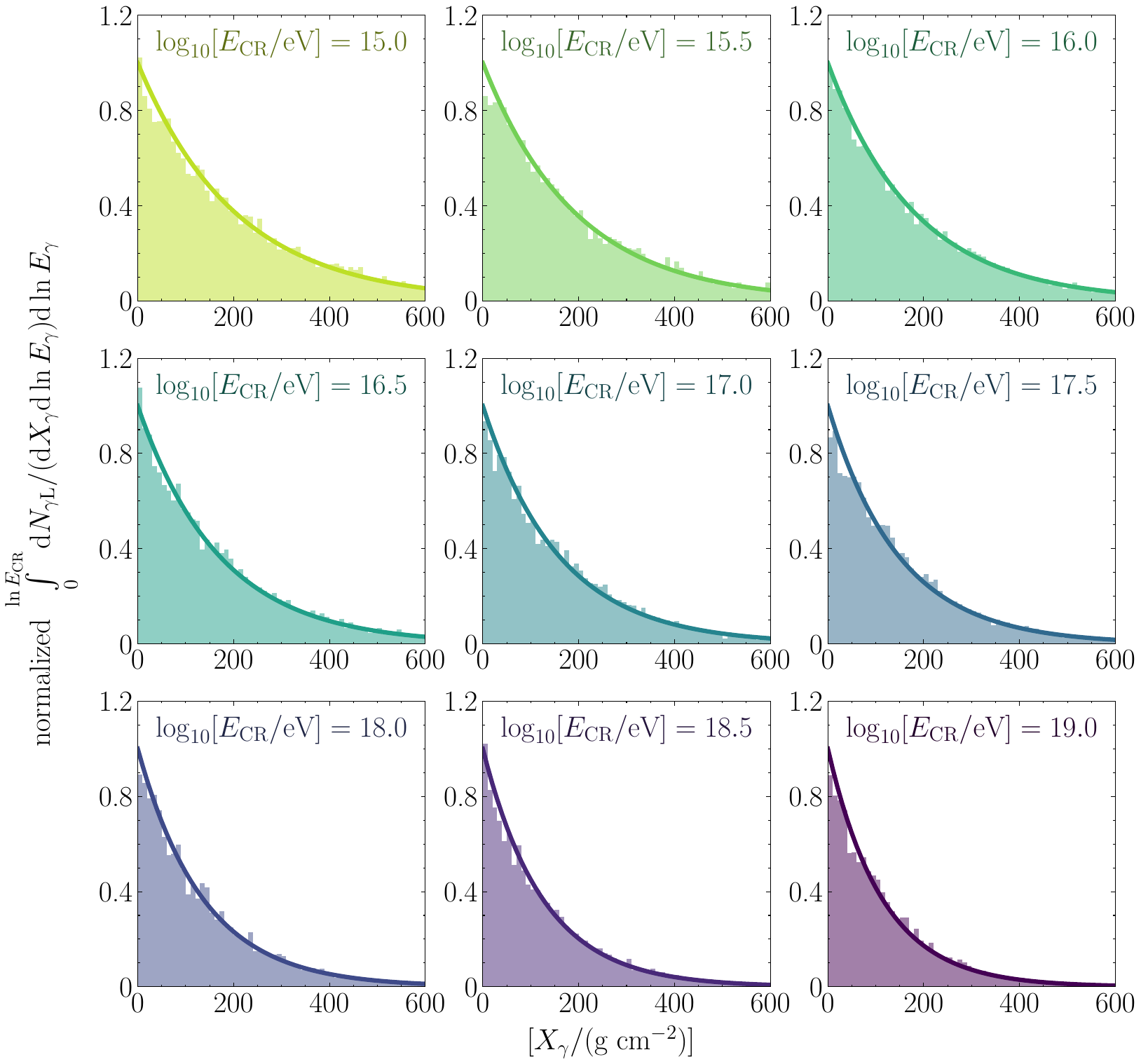}
    \caption{Best-fitting marginal hadronic photon density distributions in proton-induced EAS: \(\dd N_{\gamma\leading}/\dd \ln E_{\gamma}\) (left panel) and \(\dd N_{\gamma\leading}/\dd X_{\gamma}\) (right panel). Smooth lines represent the fitted model, while bar plots show the test dataset for different primary energies \(E_{\CR}\). For each \(E_{\CR}\), both distributions are normalized to the maximum value of the best-fitting function.}
    \label{fig:dNgamma-dlnEgammadXgamma-marginal}
\end{figure*}

The fit results are presented in Table~\ref{tab:dNgamma/dlnEgammadXgamma-parameters}. Figure~\ref{fig:dNgamma-dlnEgammadXgamma} and Figure~\ref{fig:dNgamma-dlnEgammadXgamma-marginal} compare, respectively, the best-fitting distribution of high-energy hadronic photons and the corresponding one-dimensional marginal distributions, with histograms derived from the test dataset. The comparison reveals that the maximum deviation of the best-fitting density from the empirical distribution remains within \(\sim 10\%\) of the peak value of the fitted function.

\begin{table}[htb]
    \centering
    \caption{Best-fitting parameters of \(\dd N_{\gamma\leading}/(\dd \ln E_{\gamma}\,\dd X_{\gamma})\) and their 68\% confidence level uncertainties. Energies \(E_{\CR}\) and \(E_{\gamma}\) are in eV; slant depth \(X_{\gamma}\) is in g~cm\(^{-2}\).}
    \label{tab:dNgamma/dlnEgammadXgamma-parameters}
    \vspace{\baselineskip}
    
    \begin{tabular}{l r@{\,\(\pm\)\,}l}
        \textbf{Parameter} & \multicolumn{2}{c}{\textbf{Best-fitting Value}} \\
        \hline
        \(\mathcal{I}\{\ln\mu_{[N_{\gamma\leading}]}\}\) & \(-0.486\) & \(0.116\) \\
        \(\mathcal{S}\{\ln\mu_{[N_{\gamma\leading}]},\ln E_{\CR}\}\) & \(0.163\) & \(0.003\) \\
        \(\mathcal{I}\{\mu_{[\ln E_{\gamma}]}\}\) & \(7.73\) & \(0.08\) \\
        \(\mathcal{S}\{\mu_{[\ln E_{\gamma}]},\ln E_{\CR}\}\) & \(0.489\) & \(0.002\) \\
        \(\mathcal{I}\{\varsigma_{[\ln E_{\gamma}]}\}\) & \(-3.20\) & \(0.28\) \\
        \(\mathcal{S}\{\varsigma_{[\ln E_{\gamma}]},\ln E_{\CR}\}\) & \(0.179\) & \(0.007\) \\
        \(\mathcal{I}\{\mu_{[X_{\gamma}]}\}\) & \(548\) & \(40\) \\
        \(\mathcal{S}\{\mu_{[X_{\gamma}]},\ln E_{\CR}\}\) & \(-9.92\) & \(1.00\) \\
        \(-\ln(1-\mathcal{I}\{\tilde\varsigma_{[\epsilon_{\gamma}\mid x_{\gamma}]}\})\) & \(5.74\) & \(0.15\) \\
        \(\mathcal{S}\{\tilde\varsigma_{[\epsilon_{\gamma}\mid x_{\gamma}]}, \tilde\mu_{[\epsilon_{\gamma}\mid x_{\gamma}]}\) & \(0.375\) & \(0.085\) \\
        \hline
    \end{tabular}
\end{table}

\begin{figure}[htb]
    \centering
    \includegraphics[width=\linewidth]{figs/dNgamma_dlnEgamma_dXgamma_2d.pdf}
    \caption{Best-fitting hadronic photons density distribution \(\dd N_{\gamma\leading}/(\dd \ln E_{\gamma} \dd X_{\gamma})\) in proton-induced EAS. Smooth black contours (corresponding to levels \(0.8\), \(0.6\), \(0.4\), and \(0.2\)) represent the fitted model, while colored density heatmaps show the test dataset for different primary energies \(E_{\CR}\). For each \(E_{\CR}\), both distributions are normalized to the maximum value of the best-fitting function. The corresponding normalization factors, referred to as peak densities, are shown in the plots.}
    \label{fig:dNgamma-dlnEgammadXgamma}
\end{figure}

\subsection{Photonuclear muons in the sub-showers induced by hadronic photons}
\label{subsec:analytical:photon-subshowers}

For the function \(\Gamma\) describing PN\(\mu\) production in photon-induced cascades, the following quantities are defined:
\begin{equation}
    \begin{aligned}
        &\Gamma_{\max}(E_{\gamma A}, E_{\gamma}) = \max_{X_{\mu}} \Gamma(E_{\gamma A}, E_{\gamma}, X_{\mu}),~\text{and} \\
        &X_{\mu, \max \Gamma}(E_{\gamma A}, E_{\gamma}) = \operatornamewithlimits{argmax}_{X_{\mu}} \Gamma(E_{\gamma A}, E_{\gamma}, X_{\mu}),
    \end{aligned}
\end{equation}
where \(X_{\mu, \max \Gamma}\) should not be confused with \(X_{\max}\), the position of the longitudinal profile maximum for charged particles. A rescaled slant depth is also introduced:
\begin{equation}
    x_{\mu, \Gamma} = X_{\mu,\max\Gamma}^{-1} \times X_{\mu}.
\end{equation}
The parametrization of \(\Gamma\) takes the form:
\begin{equation}
    \label{eq:Gamma-parametrization}
    \begin{aligned}
    &\Gamma(E_{\gamma A}, E_{\gamma}, X_{\mu}) = \Theta(E_{\gamma}-E_{\gamma A})\times \Theta(X_{\mu}) \times \\
    &\times \Gamma_{\max} \times \exp\left(\frac{X_{\mu,\max \Gamma}}{\Lambda_{\Gamma}}(\ln x_{\mu, \Gamma}+1-x_{\mu, \Gamma})\right),
    \end{aligned}
\end{equation}
where \(\Lambda_{\Gamma}(x_{\mu, \Gamma}, E_{\gamma A})\) denotes the effective interaction depth in the generalized Gaisser--Hillas longitudinal profile~\cite{Gaisser:1977}.

Three independent linear models are constructed for the variables \(X_{\mu, \max \Gamma}\), \(\ln \Gamma_{\max}\), and \(\Lambda_{\Gamma}\), using the following feature sets:
(i) \(\ln E_{\gamma A}\), \([\ln E_{\gamma A}]^2\), and \(\ln E_{\gamma}\) for \(\ln \Gamma_{\max}\); (ii) \(\ln E_{\gamma A}\) and \(\ln E_{\gamma}\) for \(X_{\mu, \max \Gamma}\); (iii) \(\ln E_{\gamma A}\), \(x_{\mu, \Gamma}\), \(x_{\mu, \Gamma}\ln E_{\gamma A}\), \(x_{\mu, \Gamma}^2\), and \(x_{\mu, \Gamma}^2\ln E_{\gamma A}\) for \(\Lambda_{\Gamma}\).

The training dataset is filtered to exclude outliers in the regions \(\ln E_{\gamma A} \rightarrow (\ln E_{\gamma}-0)\) and \(x_\mu \rightarrow 1\) of the feature space. The intercept and slope parameters are then fitted to the filtered training data, and their uncertainties are estimated via standard error propagation for linear regression.

\begin{figure}[htb]
    \centering
    \includegraphics[width=\linewidth]{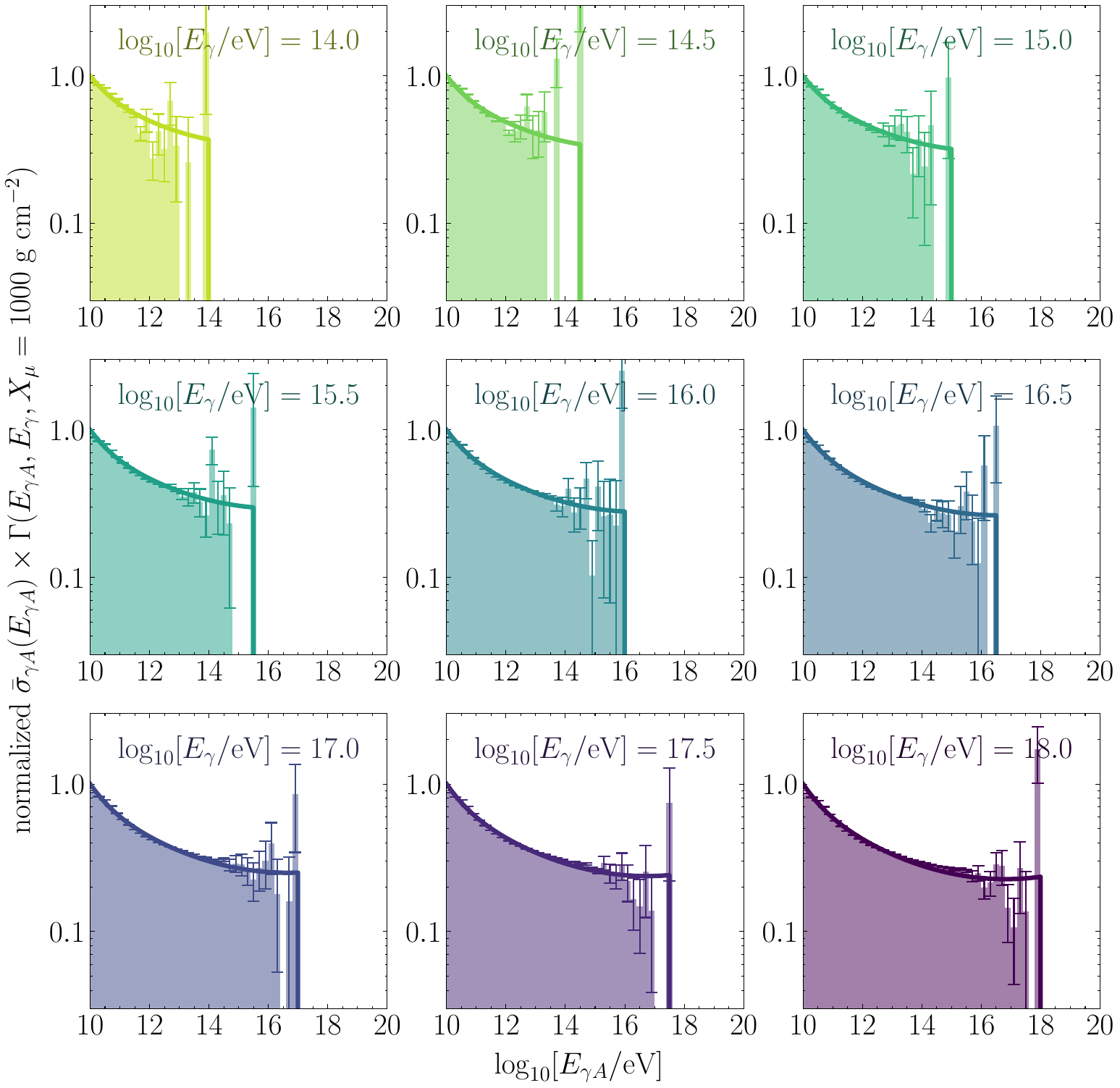}
    \caption{Best-fitting PN\(\mu\) density distribution \(\bar\xs_{\gamma A}\times \Gamma=\dd N_{\muphNucl}/\dd \ln E_{\gamma A}\) in photon-induced EAS (smooth lines) versus the test dataset (bar plots with statistical error bars) for different values of \(E_{\gamma}\) at a fixed slant depth \(X_{\mu} = 1000\,\text{g\,cm}^{-2}\). For each \(E_{\gamma}\), both distributions are normalized to the maximum of the best-fitting function.}
    \label{fig:Xsection-times-Gamma}
\end{figure}

\begin{figure}[htb]
    \centering
    \includegraphics[width=\linewidth]{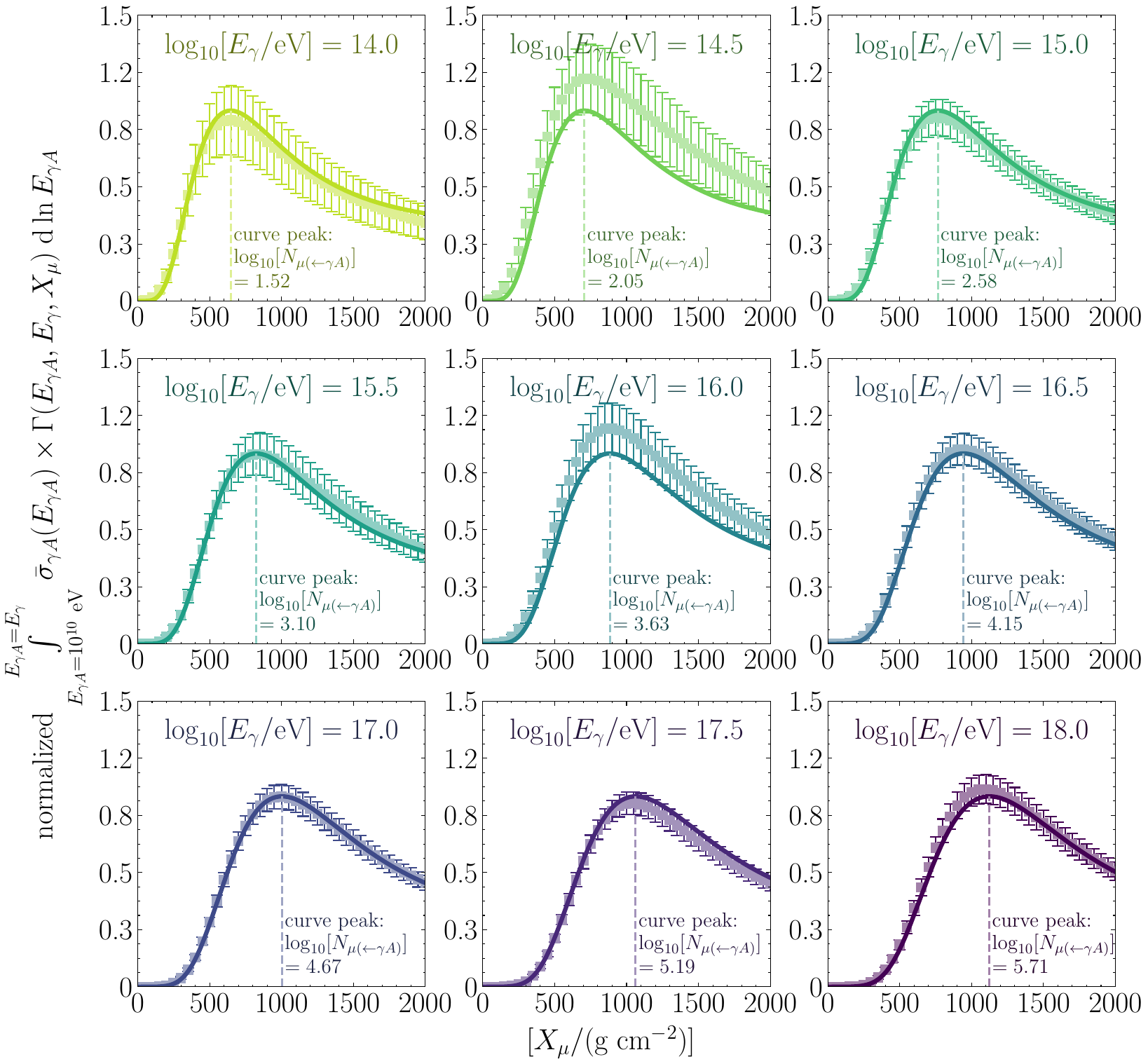}
    \caption{Longitudinal profile of the best-fitting PN\(\mu\) density distribution in photon-induced extensive air showers, integrated over the energy range \(10^{10}\,\text{eV} \leq E_{\gamma A} \leq E_{\gamma}\) (smooth lines), compared with the test dataset (points with error bars) for different \(E_{\gamma}\). For each \(E_{\gamma}\), both profiles are normalized to the maximum of the best-fitting function. The vertical dashed line indicates the position of the peak in the best‑fitting profile, and the text to the right of this line provides the normalization factor.} The test dataset is down-sampled for illustrative clarity.
    \label{fig:Xsection-times-Gamma-longi}
\end{figure}

The fit results are summarized in Table~\ref{tab:Gamma-parameters}. Figure~\ref{fig:Xsection-times-Gamma} and Figure~\ref{fig:Xsection-times-Gamma-longi} compare, respectively, the best-fitting approximation of \(\bar\xs_{\gamma A} \times \Gamma\) with values estimated from the test dataset, and the corresponding longitudinal profiles of PN\(\mu\). The model predictions agree with the test data within the statistical uncertainty of the latter.

\begin{table}[htb]
    \centering
    \caption{Best-fitting parameters of \(\Gamma\) and their 68\% confidence level uncertainties. \(\Gamma_{\max}\) is given in mb\(^{-1}\); \(X_{\mu, \max \Gamma}\) and \(\Lambda_{\Gamma}\) are in g cm\(^{-2}\); energies \(E_{\gamma A}\) and \(E_{\gamma}\) are in eV.}
    \label{tab:Gamma-parameters}
    \vspace{\baselineskip}

    \begin{tabular}{l r@{\,\(\pm\)\,}l}
        \textbf{Parameter} & \multicolumn{2}{c}{\textbf{Best-fitting Value}} \\
        \hline
        \(\mathcal{I}\{\ln\Gamma_{\max}\}\) & \(-22.7\) & \(3.5\) \\
        \(\mathcal{S}\{\ln\Gamma_{\max},\ln E_{\gamma A}\}\) & \(-0.463\) & \(0.237\) \\
        \(\mathcal{S}\{\ln\Gamma_{\max},[\ln E_{\gamma A}]^2\}\) & \((5.19 \) & \(3.98) \times 10^{-3}\) \\
        \(\mathcal{S}\{\ln\Gamma_{\max},\ln E_{\gamma}\}\) & \(1.00\) & \(0.02\) \\
        \(\mathcal{I}\{X_{\mu, \max \Gamma}\}\) & \(-1287\) & \(277\) \\
        \(\mathcal{S}\{X_{\mu, \max \Gamma},\ln E_{\gamma A}\}\) & \(16.7\) & \(7.8\) \\
        \(\mathcal{S}\{X_{\mu, \max \Gamma},\ln E_{\gamma}\}\) & \(46.2\) & \(7.9\) \\
        \(\mathcal{I}\{\Lambda_{\Gamma}\}\) & \(-244\) & \(621\) \\
        \(\mathcal{S}\{\Lambda_{\Gamma}, \ln E_{\gamma A}\}\) & \(13.4\) & \(21.3\) \\
        \(\mathcal{S}\{\Lambda_{\Gamma}, x_{\mu, \Gamma}\}\) & \(521\) & \(1377\) \\
        \(\mathcal{S}\{\Lambda_{\Gamma}, x_{\mu, \Gamma} \ln E_{\gamma A}\}\) & \(-17.6\) & \(50.2\) \\
        \(\mathcal{S}\{\Lambda_{\Gamma}, x_{\mu, \Gamma}^2\}\) & \(-261\) & \(656\) \\
        \(\mathcal{S}\{\Lambda_{\Gamma}, x_{\mu, \Gamma}^2 \ln E_{\gamma A}\}\) & \(11.9\) & \(25.0\) \\
        \hline
    \end{tabular}
\end{table}

Note that the best-fitting effective quadratic slope, \(\mathcal{S}\{\Lambda_{\Gamma}, x_{\mu, \Gamma}^2\} + \mathcal{S}\{\Lambda_{\Gamma}, x_{\mu, \Gamma}^2 \ln E_{\gamma A}\} \times \ln E_{\gamma A}\), is not positive for \(E_{\gamma A} \lesssim 3.2 \times 10^9\,\text{eV}\). This condition induces an undesired singularity where \(\Lambda_{\Gamma} = 0\) at large \(x_{\mu, \Gamma}\). Although this \(E_{\gamma A}\) region is not considered here as the “high-energy” range, regularization is required to avoid the singularity. The clipping function is defined as follows:
\begin{equation}
    \begin{aligned}
    &\operatorname{clip}(\ln E_{\gamma A}) = \max\bigl\{\ln E_{\gamma A}, \\
    &-\mathcal{S}\{\Lambda_{\Gamma}, x_{\mu, \Gamma}^2\} \,/\, \mathcal{S}\{\Lambda_{\Gamma}, x_{\mu, \Gamma}^2 \ln E_{\gamma A}\}\bigr\},
    \end{aligned}
\end{equation}
and the singular \(\Lambda_{\Gamma}(x_{\mu, \Gamma}, \ln E_{\gamma A})\) is subsequently replaced with its regularized counterpart \(\Lambda_{\Gamma}(x_{\mu, \Gamma}, \operatorname{clip}(\ln E_{\gamma A}))\), for numerical robustness.

\subsection{Photonuclear muon content of proton-induced extensive air showers}
\label{subsec:analytical:joint}

To evaluate \(G_{\leading}\), the best-fitting parameterizations from Eq.~\eqref{eq:dNgamma/dlnEgammadXgamma-parametrization} and Eq.~\eqref{eq:Gamma-parametrization} are substituted into Eq.~\eqref{eq:G-factorization}.

A \(91 \times 41 \times 201\) three-dimensional grid is then introduced, spanning the relevant region of the \((E_{\gamma A},\,E_{\CR},\,X_{\mu})\) space. The grid vertices correspond to: (i) the incident photon energy range \(10^{10}\,\text{eV} \leq E_{\gamma A} \leq 10^{19}\,\text{eV}\); (ii) the primary proton energy range \(10^{15}\,\text{eV} \leq E_{\CR} \leq 10^{19}\,\text{eV}\); (iii) the slant depth range \(0 \leq X_{\mu} \leq 2000\,\text{g\,cm}^{-2}\). The spacing is \(0.1\)~dex for the energies and \(10\,\text{g\,cm}^{-2}\) for the slant depth.

The double integral is numerically computed at each grid vertex using \texttt{scipy.integrate.dblquad}, which implements QUADPACK subroutines~\cite{Piessens:2012}. The relative tolerance of the inner one-dimensional integrals is fixed at \(\tau_{\text{rel}} = 10^{-4}\), and the absolute tolerance at \(\tau_{\text{abs}} = 10^{-15}\). Once either \(\tau_{\text{rel}} = 10^{-4}\) or \(\tau_{\text{abs}} = 10^{-15}\) is reached, the algorithm terminates and disregards the remaining tolerance criterion (\(\tau_{\text{abs}}\) or \(\tau_{\text{rel}}\), respectively). In nearly all cases, except in regions where the integration domain is very small, \(\tau_{\text{abs}} = 10^{-15}\) represents an overly strict requirement, so the relative tolerance parameter dominates. This ensures that the numerical integration accuracy far exceeds the characteristic uncertainty of the integrand, which originates from the fit error.  

Analogously to Subsection~\ref{subsec:analytical:photon-subshowers}, for the function \(G_{\leading}\), the following quantities are defined:
\begin{equation}
    \begin{aligned}
        &G_{\max}(E_{\gamma A}, E_{\CR}) = \max_{X_{\mu}} G_{\leading}(E_{\gamma A}, E_{\CR}, X_{\mu}), \\
        &X_{\mu, \max G}(E_{\gamma A}, E_{\CR}) = \operatornamewithlimits{argmax}_{X_{\mu}} G_{\leading}(E_{\gamma A}, E_{\CR}, X_{\mu}),
    \end{aligned}
\end{equation}
and the corresponding rescaled (dimensionless) slant depth is introduced:
\begin{equation}
    x_{\mu, G} = X_{\mu,\max G}^{-1} \times X_{\mu}.
\end{equation}
It should be emphasized again that \(X_{\mu,\max G}\) is unrelated to the conventional \(X_{\max}\) (longitudinal profile maximum for charged particles).

A parametrization of \(G_{\leading}\) is then introduced, similar to that of \(\Gamma\):
\begin{equation}
    \label{eq:G-parametrization}
    \begin{aligned}
    &G_{\leading}(E_{\gamma A}, E_{\CR}, X_{\mu}) = \Theta(E_{\CR}-E_{\gamma A})\times\Theta(X_{\mu}) \times \\
    &\times G_{\max} \times \exp\left(\frac{X_{\mu,\max G}}{\Lambda_{G}}(\ln x_{\mu, G}+1-x_{\mu, G})\right),
    \end{aligned}
\end{equation}
where \(\Lambda_{G}(x_{\mu, G}, E_{\gamma A})\) denotes the effective interaction depth in the generalized Gaisser--Hillas longitudinal profile~\cite{Gaisser:1977}.

The quantity \(G_{\max}\) is parametrized as follows:
\begin{equation}
    G_{\max} = G_{\max}^{\text{lin}} \times \operatorname{cutoff}\left(\frac{E_{\gamma A}}{E_{\CR}},\mathcal{R},\mathcal{P}\right),
\end{equation}
where \(\ln G_{\max}^{\text{lin}}\) is linear with respect to \(\ln E_{\gamma A}\) and \(\ln E_{\CR}\), \(\mathcal{R}>0\) and \(\mathcal{P}>0\) are parameters, and
\begin{equation}
    \operatorname{cutoff}(\square,\mathcal{R},\mathcal{P}) = 
    \begin{cases} 
        1 - \left(\frac{\square}{\mathcal{R}}\right)^{\mathcal{P}}, & \square \leq \mathcal{R}, \\
        0, & \square \geq \mathcal{R}.
    \end{cases}
\end{equation}

The parameters of \(G_{\max}\) are optimized by solving the corresponding nonlinear least-squares problem for \(\ln G_{\max}\) using \texttt{scipy.optimize.curve\_fit}, which implements the Levenberg--Marquardt algorithm from MINPACK subroutines~\cite{More:1978}. For numerical stability, \((-\ln \mathcal{R})\) is optimized instead of \(\mathcal{R}\).

The fitting procedure for \(X_{\mu,\max G}\) and \(\Lambda_{G}\) follows the same approach as that for \(X_{\mu,\max\Gamma}\) and \(\Lambda_{\Gamma}\), with \(E_{\gamma}\) and \(x_{\mu,\Gamma}\) replaced by \(E_{\CR}\) and \(x_{\mu,G}\), respectively (see Subsection~\ref{subsec:analytical:photon-subshowers}). For \(\Lambda_G\), an analogous regularization technique is applied -- specifically, clipping \(\ln E_{\gamma A}\) in the region \(E_{\gamma A} \lesssim 5.9 \times 10^8\,\text{eV}\).

\begin{figure}[htb]
    \centering
    \includegraphics[width=\linewidth]{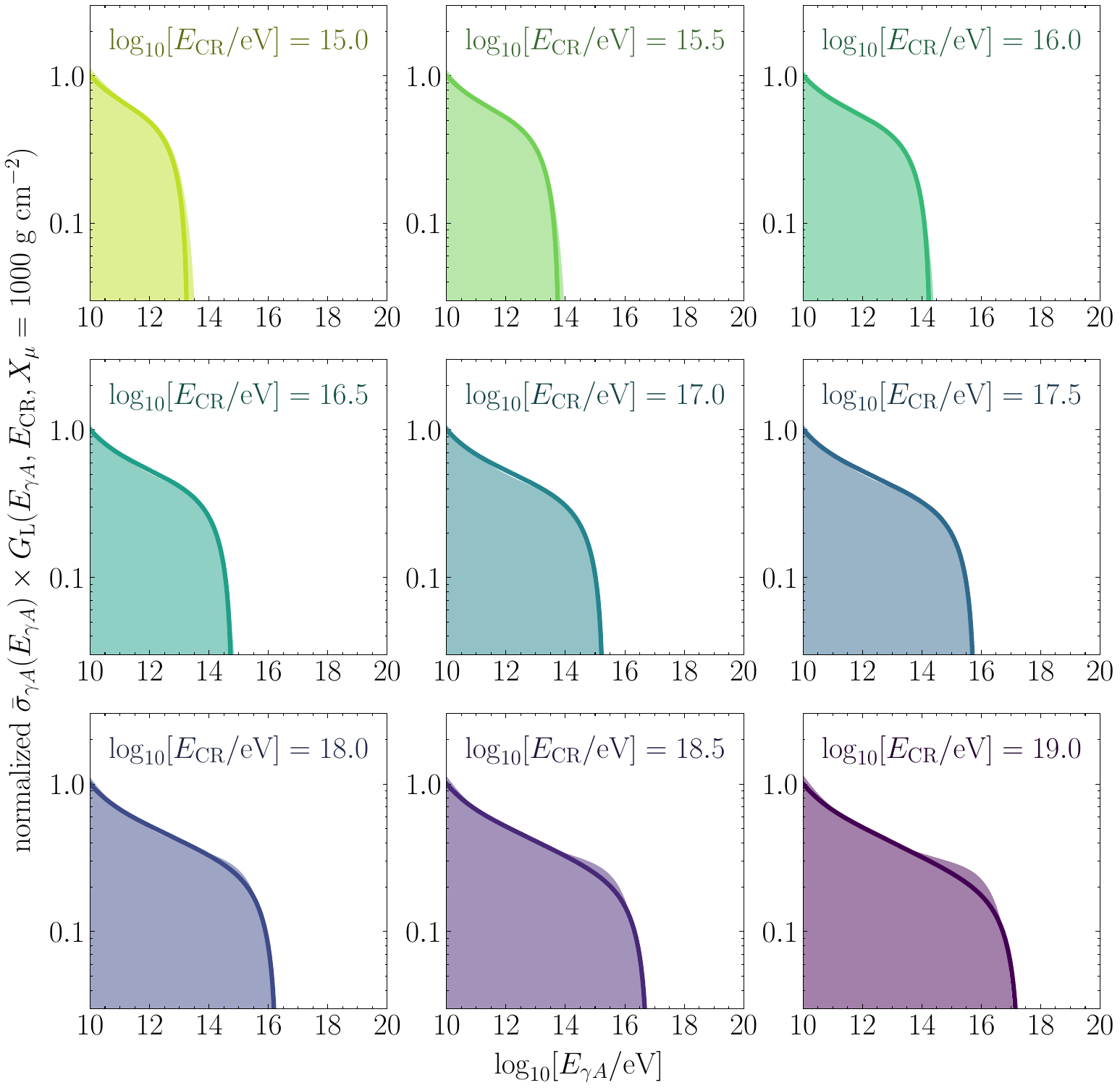}
    \caption{Best-fitting PN\(\mu\) density distribution \(\bar\xs_{\gamma A}\times G_{\leading} = \dd N_{\muphNucl}/\dd \ln E_{\gamma A}\) in proton-induced EAS (smooth lines) versus the distribution estimated via numerical integration (shaded region) for different values of \(E_{\CR}\) at a fixed slant depth \(X_{\mu} = 1000\,\text{g\,cm}^{-2}\). For each \(E_{\CR}\), both distributions are normalized to the maximum of the best-fitting function.}
    \label{fig:Xsection-times-G}
\end{figure}

\begin{figure}[htb]
    \centering
    \includegraphics[width=\linewidth]{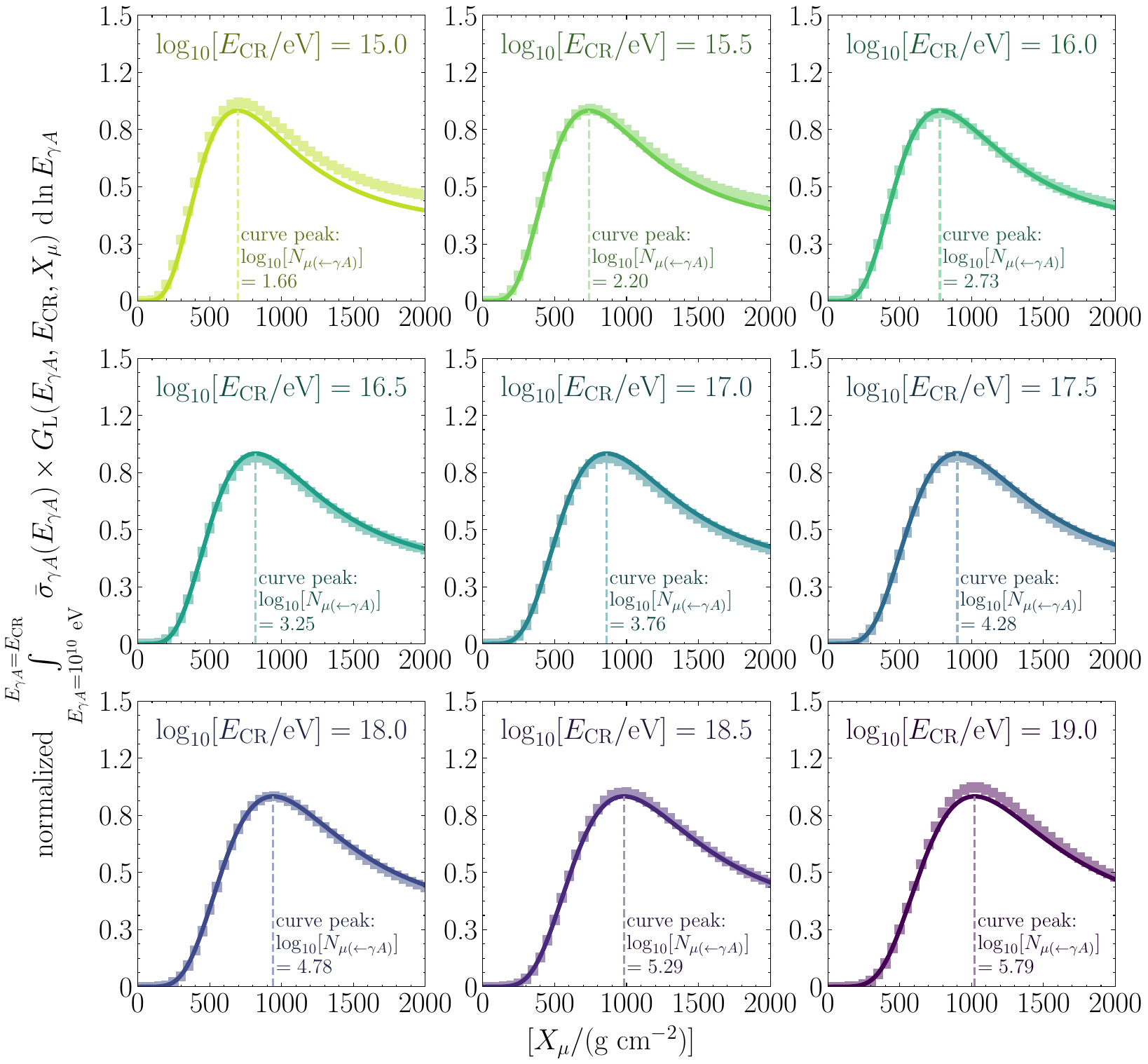}
    \caption{Longitudinal profile of the best-fitting PN\(\mu\) density distribution in proton-induced EAS, integrated over the energy range \(10^{10}\,\text{eV} \leq E_{\gamma A} \leq E_{\CR}\) (smooth lines), compared with the profile estimated via numerical integration (points) for different \(E_{\CR}\). For each \(E_{\CR}\), both profiles are normalized to the maximum of the best-fitting function. The vertical dashed line indicates the position of the peak in the best‑fitting profile, and the text to the right of this line provides the normalization factor.} The data points are down-sampled for illustrative clarity.
    \label{fig:Xsection-times-G-longi}
\end{figure}

The fit results are summarized in Table~\ref{tab:G-parameters}. Figure~\ref{fig:Xsection-times-G} and Figure~\ref{fig:Xsection-times-G-longi} compare, respectively, the best-fitting approximation of \(\bar\xs_{\gamma A} \times G_{\leading}\) with values estimated via numerical integration and the corresponding longitudinal profiles of PN\(\mu\). The comparison demonstrates that the best-fitting model defined by Eq.~\eqref{eq:G-parametrization} provides an acceptable approximation of the integral given by Eq.~\eqref{eq:G-factorization}.

\begin{table}[htb]
    \centering
    \caption{Best-fitting parameters of \(G_{\leading}\) and their 68\% confidence level uncertainties. \(G^{\text{lin}}_{\max}\) is given in mb\(^{-1}\); \(X_{\mu, \max G}\) and \(\Lambda_{G}\) are in g\,cm\(^{-2}\); energies \(E_{\gamma A}\) and \(E_{\CR}\) are in eV.}
    \label{tab:G-parameters}
    \vspace{\baselineskip}

    \begin{tabular}{l r@{\,\(\pm\)\,}l}
        \textbf{Parameter} & \multicolumn{2}{c}{\textbf{Best-fitting Value}} \\
        \hline
        \(\mathcal{I}\{\ln G^{\text{lin}}_{\max}\}\) & \(-27.71\) & \(0.05\) \\
        \(\mathcal{S}\{\ln G^{\text{lin}}_{\max},\ln E_{\gamma A}\}\) & \(-0.166\) & \(0.001\) \\
        \(\mathcal{S}\{\ln G^{\text{lin}}_{\max},\ln E_{\CR}\}\) & \(0.976\) & \(0.002\) \\
        \(-\ln \mathcal{R}\) & \(3.94\) & \(0.01\) \\
        \(\mathcal{P}\) & \(0.683\) & \(0.020\) \\
        \(\mathcal{I}\{X_{\mu, \max G}\}\) & \(-778\) & \(162\) \\
        \(\mathcal{S}\{X_{\mu, \max G},\ln E_{\gamma A}\}\) & \(16.7\) & \(2.7\) \\
        \(\mathcal{S}\{X_{\mu, \max G},\ln E_{\CR}\}\) & \(30.2\) & \(4.2\) \\
        \(\mathcal{I}\{\Lambda_{G}\}\) & \(-154\) & \(337\) \\
        \(\mathcal{S}\{\Lambda_{G}, \ln E_{\gamma A}\}\) & \(10.7\) & \(11.5\) \\
        \(\mathcal{S}\{\Lambda_{G}, x_{\mu, G}\}\) & \(464\) & \(711\) \\
        \(\mathcal{S}\{\Lambda_{G}, x_{\mu, G} \ln E_{\gamma A}\}\) & \(-16.7\) & \(25.3\) \\
        \(\mathcal{S}\{\Lambda_{G}, x_{\mu, G}^2\}\) & \(-246\) & \(316\) \\
        \(\mathcal{S}\{\Lambda_{G}, x_{\mu, G}^2 \ln E_{\gamma A}\}\) & \(12.2\) & \(11.6\) \\
        \hline
    \end{tabular}
\end{table}

\section{Results and Discussion}
\label{sec:disc}
This section quantifies the quality of the best-fitting model over a wide range of parameter space, including different PNR cross section models, primary particle types and energies, as well as different slant depths. In addition, the key limitations are discussed, and the best‑fitting model is employed to characterize the PN\(\mu\) content, assuming the commonly used EGS4 model for the PNR cross section.

\subsection{Relative error distribution}
The quantity \(\Delta N_\mu\) is defined as the average (over a large number of events) difference in the number of EAS muons between simulations using the modified PNR cross section and the standard EGS4 cross section. According to the model, this difference is only due to the PN\(\mu\), \(\Delta N_{\mu} = \Delta N_{\muphNucl}\). Therefore, \(\Delta N_{\mu}\) can be estimated by integrating the product of the best-fitting Green’s Function \(\delta N_{\muphNucl}/\delta\bar{\xs}_{\gamma A}\) and the difference of the corresponding cross sections, \(\Delta\bar{\xs}_{\gamma A}\), as shown in Eq.~\eqref{eq:linear-functional}. This estimation is hereafter referred to as \(\Delta \hat{N}_{\mu}\).

For the family of PNR cross sections defined by Eq.~\eqref{eq:parametricXsection}, \(\Delta\bar{\xs}_{\gamma A}\) can be expressed as
\begin{equation}
    \label{eq:DeltaXsection}
    \begin{aligned}
        \Delta\bar{\xs}_{\gamma A}(E_{\gamma A} \mid E_{\gamma A}^{\thr}, a) = \bar{\xs}_{\gamma A}(E_{\gamma A} \mid \text{EGS4}) \times \\
        \times \begin{cases}
            (a-1) \times \operatorname{smooth}\left(\frac{E_{\gamma A}}{E_{\gamma A}^{\thr}}\right), & E_{\gamma A} \geq E_{\gamma A}^{\thr}, \\
            0, & E_{\gamma A} < E_{\gamma A}^{\thr}.
        \end{cases}
    \end{aligned}
\end{equation}
Therefore, \(\Delta \hat{N}_{\mu}\) can be easily computed using numerical integration. The \texttt{scipy.integrate.quad} function is used here, which implements QUADPACK subroutines~\cite{Piessens:2012}.

On the other hand, \(\Delta N_{\mu}\) can be estimated directly from MC simulations. The simulations discussed in Subsection~\ref{subsec:monte-carlo:modified-cs} are used here, and this estimation is hereafter referred to as \(\Delta N_{\mu}^{\text{MC}}\).

To evaluate the performance, the relative error metric is defined as
\begin{equation}
    \label{eq:DeltaNmu-rel-err}
    \operatorname{rel.\,err.}\,\{\Delta N_{\mu}\} = \frac{\Delta \hat{N}_{\mu} - \Delta N_{\mu}^{\text{MC}}}{\Delta N_{\mu}^{\text{MC}}},
\end{equation}
and this metric is computed for various available combinations of parameters.

\begin{figure}[htb]
    \centering
    \includegraphics[width=\linewidth]{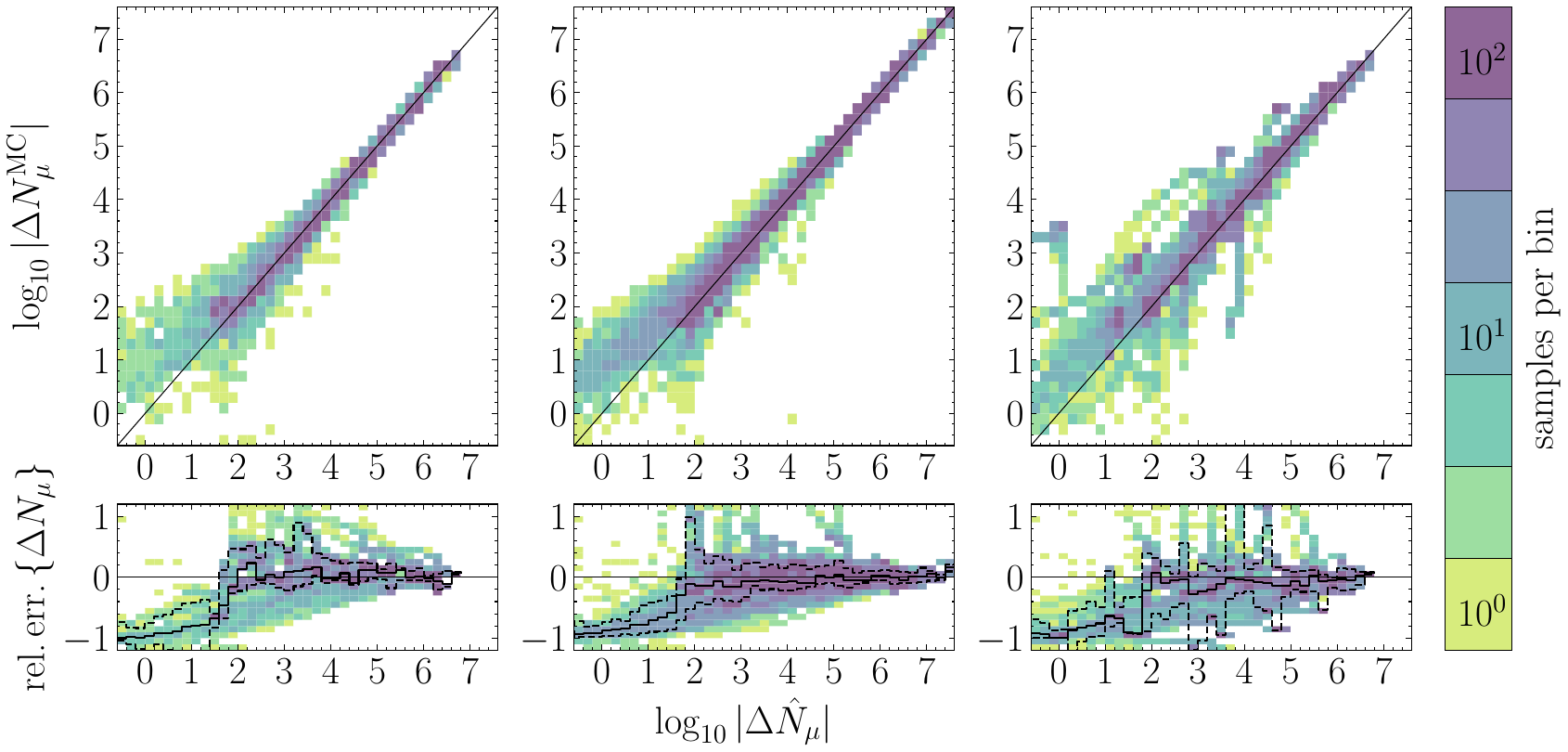}
    \caption{Density distributions of the MC result (top) and the relative error of the best-fitting model prediction (bottom) against the prediction \(\Delta \hat{N}_{\mu}\). 
    Left: different \(E_{\CR(\gamma)}\), \(A_{\CR}\), and \(X_{\mu}\), with common \(E_{\gamma A}^{\thr} = 10^{10}\,\text{eV}\) and \(a = 2\).
    Center: different \(E_{\gamma,\,\CR}\), \(X_{\mu}\), and \(a\), with common \(E_{\gamma A}^{\thr} = 10^{10}\,\text{eV}\), only photon- and proton-induced EAS.
    Right: different \(E_{\gamma,\,\CR}\), \(X_{\mu}\), and \(E_{\gamma A}^{\thr}\), with common \(a = 2\), only photon- and proton-induced EAS.
    The color represents the number of unique parameter combinations \((E_{\gamma A}^{\thr},\,a,\,E_{\CR},\,A_{\CR},\,X_{\mu})\) (or \((E_{\gamma A}^{\thr},\,a,\,E_{\gamma},\,X_{\mu})\) in the case of photon-induced EAS) per bin.
    The solid black lines on the bottom plots correspond to the conditional median of \(\operatorname{rel.\,err.}\,\{\Delta N_{\mu}\}\), while the dashed black lines represent the edges of its conditional symmetric \(68\%\) confidence level interval.}
    \label{fig:val-joint}
\end{figure}

Figure~\ref{fig:val-joint} summarizes the relative error metric distribution for all simulations described in Subsection~\ref{subsec:monte-carlo:modified-cs} (note that these MC data sets intersect at \(a=2\), \(E_{\gamma A} = 10^{10}\)~eV). In general, an acceptable agreement between the MC data and the best-fitting model predictions can be observed across a wide region of parameter space. For large \(\Delta N_{\mu} \gtrsim 10^5\) (relevant in the context of the muon puzzle), the characteristic relative error is about \((10\dots 20)\%\), while for small \(\Delta N_{\mu} \lesssim 10^3\), the model tends to underestimate the difference in muon numbers. The latter, however, does not reduce the usefulness of the model, since the prediction remains conservative, and the actual effect is too small to be significant in the context of the muon puzzle (or even to be reliably observed).

\begin{figure}[htb]
    \centering
    \includegraphics[width=\linewidth]{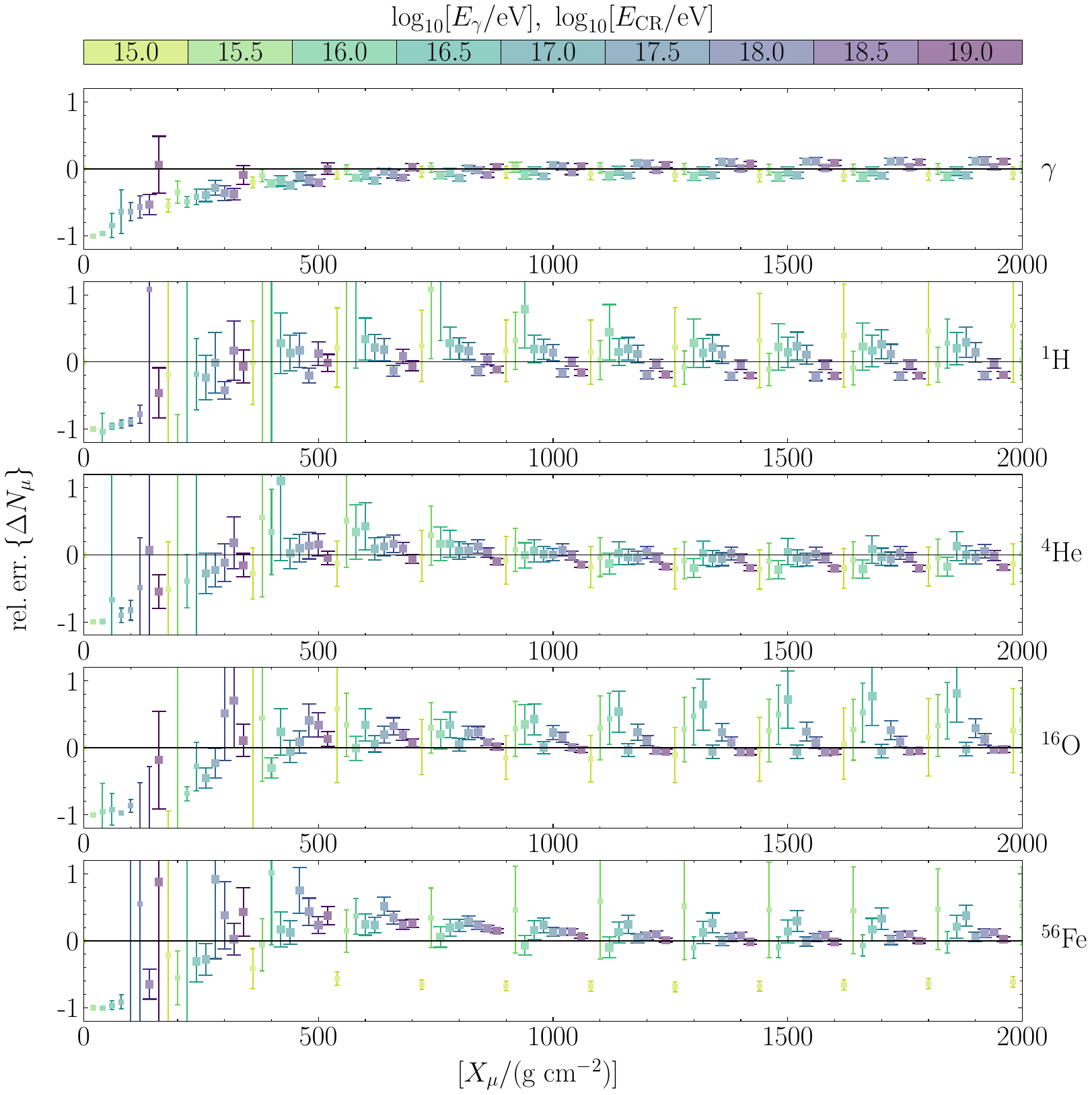}
    \caption{Longitudinal profiles of \(\operatorname{rel.\,err.}\,\{\Delta N_{\mu}\}\) for \(a = 2\), \(E_{\gamma A}^{\thr} = 10^{10}\,\text{eV}\), and for different energies and types of primary particles. The data points are down-sampled for illustrative clarity. The small markers correspond to relatively small \(\Delta N_{\mu}^{\text{MC}} < 10^3\).}
    \label{fig:val-composition-detailed}
\end{figure}

Figure~\ref{fig:val-composition-detailed} demonstrates the longitudinal profiles of the relative error metric for different types and energies of primary particles at a fixed PNR cross section model. An acceptable agreement can again be observed at the slant depths of \(X_{\mu} \gtrsim 1000\,\text{g\,cm}^{-2}\) and across the entire range of primary particle types and energies. In particular, this agreement justifies the Superposition model assumption as well as the choice of the fudge factor \(\kappa_G = 5\).

\begin{figure*}[htb]
    \centering
    \includegraphics[width=0.7\linewidth]{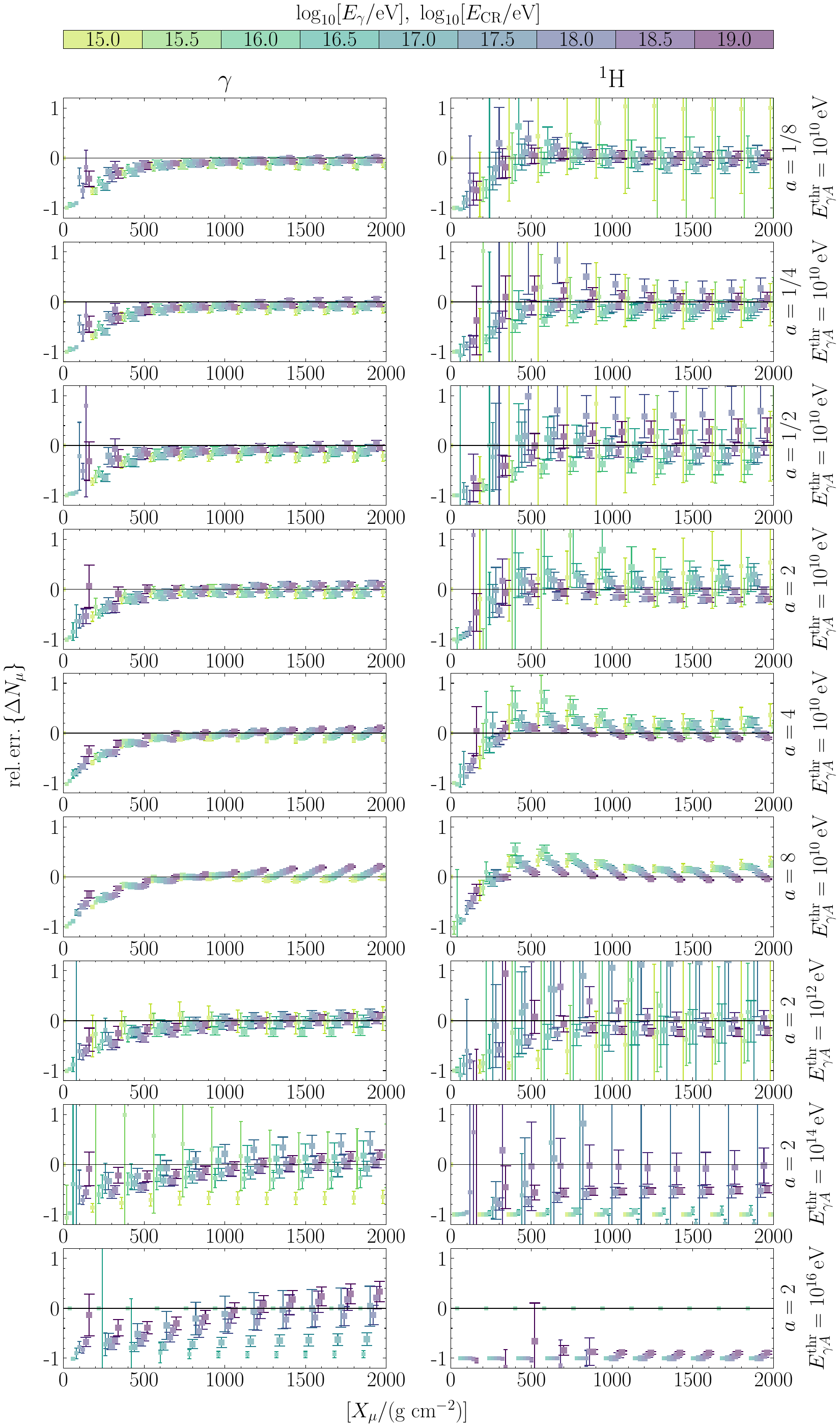}
    \caption{Longitudinal profiles of \(\operatorname{rel.\,err.}\,\{\Delta N_{\mu}\}\) in photon-induced (left) and proton-induced (right) EAS for different PNR cross section parameters \(a\) and \(E_{\gamma A}^{\thr}\), and for different energies of primary particles. The data points are down-sampled for illustrative clarity. The trivial results for \(E_{\gamma A}^{\thr}>E_{\gamma}\), \(E_{\gamma A}^{\thr} > E_{\CR}\) are omitted.} The small markers correspond to relatively small \(\Delta N_{\mu}^{\text{MC}} < 10^3\).
    \label{fig:val-amplitude-threshold-detailed}
\end{figure*}

Figure~\ref{fig:val-amplitude-threshold-detailed} shows the longitudinal profiles of the relative error metric for different PNR cross sections and different energies of primary particles in photon-induced and proton-induced EAS. The observed agreement between the MC data and the predictions of the best-fitting model evidences the applicability of the Green’s Function method across a wide range of PNR cross sections.

Note that the largest tension between the MC data and the predictions of the best-fitting model corresponds to relatively small slant depths, where the EAS are generally underdeveloped, or to the thresholds \(E_{\gamma A}^{\thr}\) near the cutoff of \(G_{\leading}\) (or \(\Gamma\), in the case of photon-induced EAS). Thus, these discrepancies are merely a manifestation of the prediction bias at small \(\Delta N_{\mu}\), as discussed above. It should be noted again that, despite this uncertainty, the best-fitting model still provides a conservative estimate, i.e., a reliable lower bound for the number of PN\(\mu\).

\subsection{Limitations}
In addition to the obvious precautions regarding extrapolation of the method beyond its fitting domains, the approach imposes two important restrictions on the high-energy PNR cross section. These limitations concern: (i) the magnitude of \(\bar{\xs}_{\gamma A}\); (ii) the high-energy asymptotic behavior of \(\dd \bar{\xs}_{\gamma A}/\dd \ln E_{\gamma A}\).

The first limitation arises from the linearity assumption. The methodology implicitly assumes a single (grand-...)parent high-energy PNR in the PN\(\mu\) interaction history. However, a large PNR cross section (comparable to that of \(e^+e^-\) pair production) can significantly distort the development of electromagnetic sub-cascades and convert a considerable amount of their energy back into hadronic cascades, enabling multiple successive high-energy PNRs.

Although the linear regime applies to a wide range of PNR cross section models, it requires the ratio \(\Delta \bar{\xs}_{\gamma A}(E_{\gamma A}) / \bar{\xs}_{\gamma A}(E_{\gamma A} \mid \text{EGS4})\) to remain much smaller than the inverse of the branching ratio for PNRs in EGS4 (\(\sim 10^{2}\)). However, even in the nonlinear regime, the proposed linear model’s predictions may still provide a useful and conservative lower bound for the number of muons.

The second limitation has a numerical origin. Near the cutoff energy, MC simulation data exhibit large fluctuations, preventing the proposed heuristic parameterization from accurately capturing the actual behavior of the Green’s Function within this region. Consequently, if the contribution of this region to the integral in Eq.~\eqref{eq:linear-functional} is substantial, the prediction will have a high degree of uncertainty. This imposes an upper limit on \(\dd \ln \bar{\xs}_{\gamma A}/\dd \ln E_{\gamma A}\) at large \(E_{\gamma A}\).

The limit is approximately \(-\mathcal{S}\{\ln G^{\text{lin}}_{\max}, \ln E_{\gamma A}\}\) (\(-\mathcal{S}\{\ln\Gamma_{\max}, \ln E_{\gamma A}\}\)) for nucleus-induced (photon-induced) EAS, i.e. \(0.166\) and \(0.463\), respectively. For comparison, in EGS4, \(\dd \ln \bar{\xs}_{\gamma A}/\dd \ln E_{\gamma A} \simeq 0.073\), see Chapter~7 of ref.~\cite{Heck:1998}.

\subsection{Muon content of extensive air showers}

Once an acceptable quality of the best-fitting model has been validated, the model can be employed in its own fitting domain to study the  muon content of EAS assuming the commonly used EGS4 model for PNRs.

\begin{figure}[htb]
    \centering
    \includegraphics[width=\linewidth]{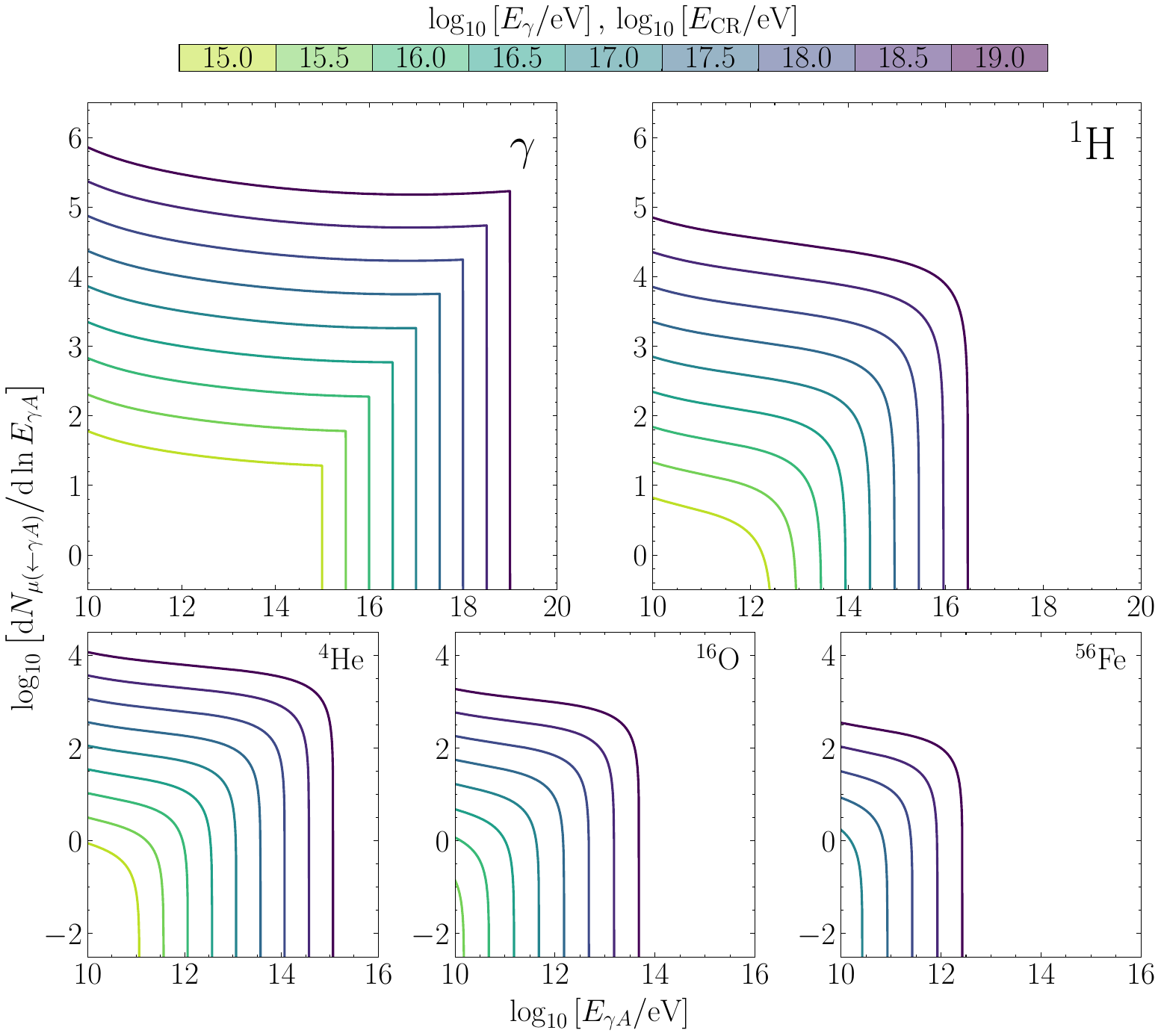}
    \caption{Best-fitting PN\(\mu\) density distribution \(\dd N_{\muphNucl}/\dd \ln E_{\gamma A}\) for different types and energies of primary particles at a fixed slant depth \(X_{\mu}=1000\,\text{g cm}^{-2}\), assuming the EGS4 PNR cross section.}
    \label{fig:PhotonuclearMuons}
\end{figure}

Figure~\ref{fig:PhotonuclearMuons} highlights the primary energy range of PNRs' incident photons that contribute most to the PN\(\mu\) yield by demonstrating the best‑fitting PN\(\mu\) density distributions \(\dd N_{\muphNucl}/\dd \ln E_{\gamma A}\) for different EAS.

This density distribution function illustrates the sensitivity of the total PN\(\mu\) yield to relative variations in the PNR cross section, \(\Delta \bar{\sigma}_{\gamma A}(E_{\gamma A}) / \bar{\sigma}_{\gamma A}(E_{\gamma A} \mid \text{EGS4})\). Notably, the density distribution exhibits a near‑plateau behavior in the region below the cutoff, making the cross‑section variations across this \(E_{\gamma A}\) domain roughly equally important. In contrast, for energies \(E_{\gamma A} \gtrsim 10^{18}\,\text{eV}\), these variations have no impact on nucleus‑induced EAS, even at the highest primary energies \(E_{\text{CR}}\), because these EAS lack sufficiently energetic photons. However, the PN\(\mu\) content of photon‑induced EAS is highly sensitive to these variations.

Therefore, a hypothetical increase in the PNR cross section at the highest energies would leave no observable signature in the muon content of nucleus-induced EAS. However, it would make photon-induced EAS more muon-rich, causing them to mimic nucleus-induced events. This introduces a considerable degree of uncertainty in the search for ultra‑high‑energy photons (\(E_{\gamma} \gtrsim 10^{18}\,\text{eV}\)) via EAS observations. A heavier primary cosmic-ray composition implies a lower cutoff in \(E_{\gamma A}\), compounding the problem. Moreover, the different sensitivity to variations in the PNR cross section between heavy and light nuclei may obstruct the reconstruction of the cosmic-ray composition itself.

\begin{figure*}[p]
    \centering
    \includegraphics[height=0.7\linewidth,angle=270]{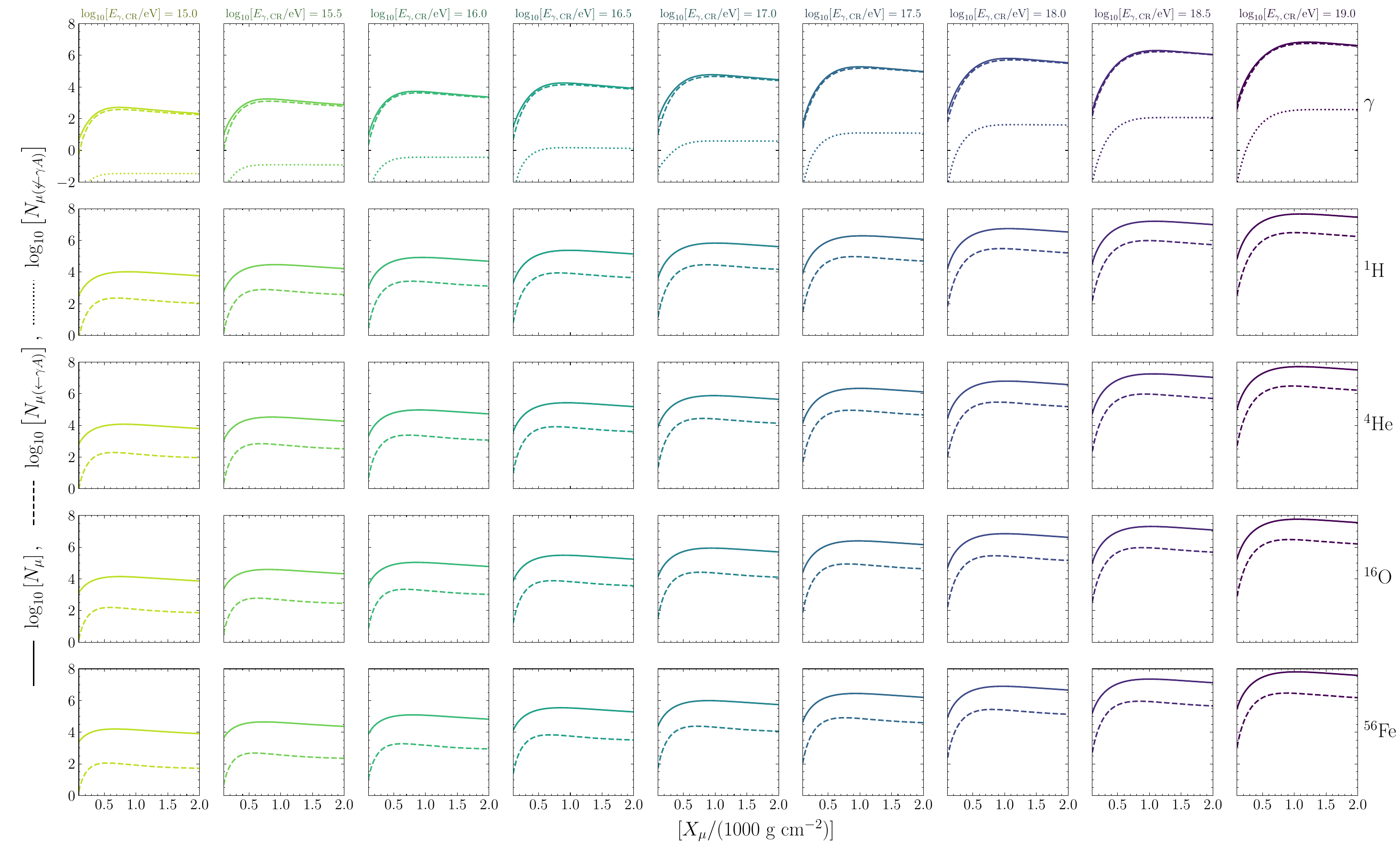}
    \caption{Longitudinal profiles of \(N_{\mu(\leftarrow \gamma A)}\) (best‑fitting model, dashed lines) and \(N_{\mu}\) (MC mean, solid lines) for a fixed threshold energy \(E_{\gamma A}^{\thr} = 10^{10}\,\text{eV}\), and for different energies and types of primary particles, assuming the EGS4 model for the PNR cross section. For photon-induced EAS, the MC mean longitudinal profiles of \(N_{\mu}\) in the case \(\bar{\sigma}_{\gamma A}(E_{\gamma A}) = 0\) (referred to as \(N_{\mu(\not\leftarrow\gamma A)}\), dotted lines) are also shown. The region \(X_{\mu} < 100\,\text{g}\,\text{cm}^{-2}\), where EAS are generally underdeveloped and best‑fitting model predictions are uncertain, is excluded from the analysis.}
    \label{fig:comparison}
\end{figure*}

\begin{figure*}[p]
    \centering
    \includegraphics[height=0.7\linewidth,angle=270]{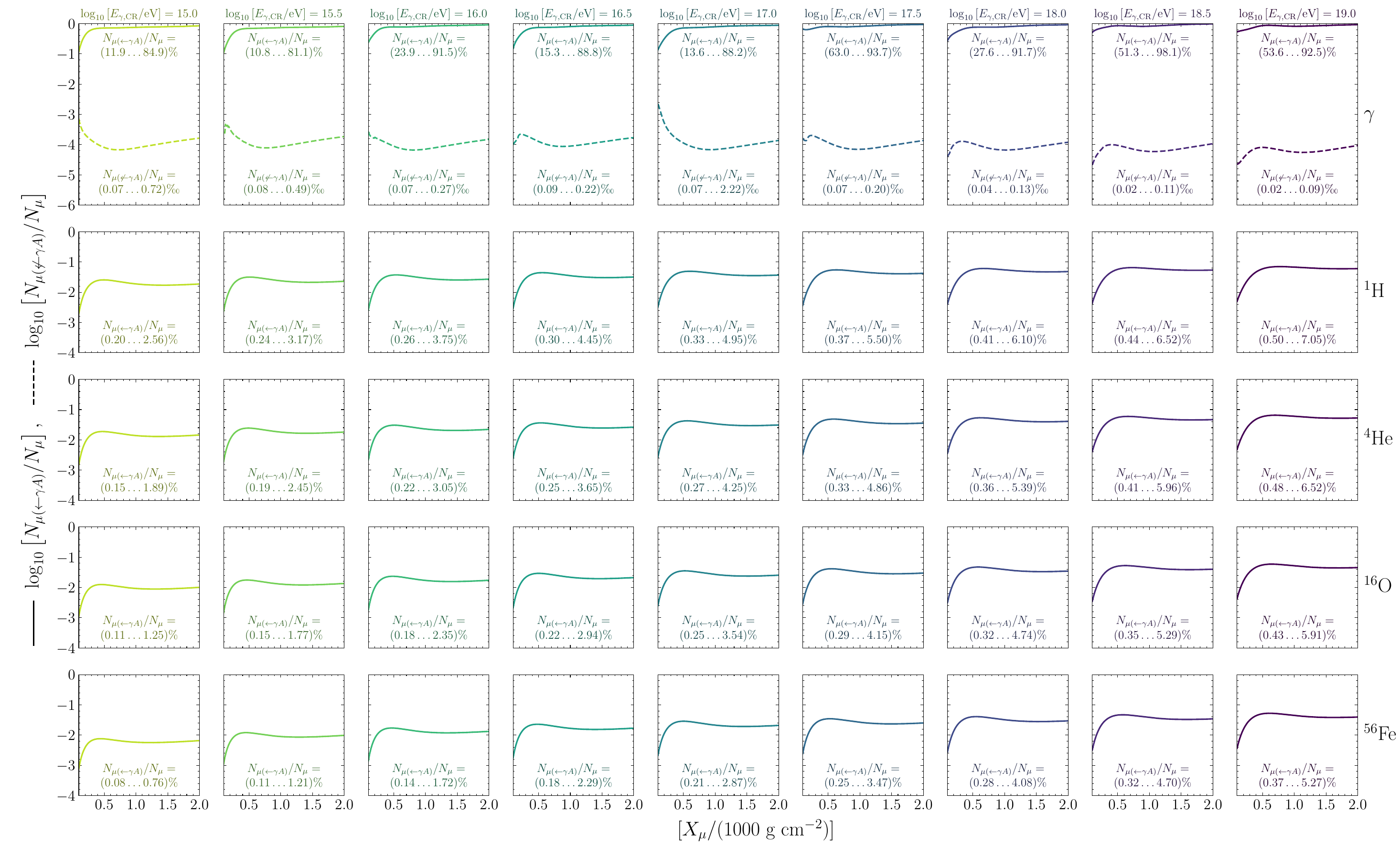}
    \caption{Same as Figure~\ref{fig:comparison}, but showing the ratios of \(N_{\mu(\leftarrow \gamma A)}\) (solid lines) and \(N_{\mu(\not\leftarrow\gamma A)}\) (dashed lines; shown only for photon-induced EAS) to \(N_{\mu}\), rather than the absolute number of muons.}
    \label{fig:ratio}
\end{figure*}

Figure~\ref{fig:comparison} compares the longitudinal profiles of the total number of muons and the PN\(\mu\) component for different primary particle types and energies, again assuming the EGS4 model. For photon‑induced EAS, the longitudinal profiles with PNR reactions disabled (\(\bar\sigma_{\gamma A}=0\)) are also shown; the corresponding number of muons is denoted as \(N_{\mu(\not\leftarrow \gamma A)}\). Figure~\ref{fig:ratio} shows the corresponding ratios of the number of muons instead of directly comparing the profiles.

It can be seen that PN\(\mu\) are the dominant part of the photon‑induced EAS muon content at the ground level. The typical fraction of the non‑photonuclear muons does not exceed \(\sim 10^{-3}\), hence the remaining part of the muons is produced via photonuclear reactions in the resonant region (\(E_{\gamma A}<10^{10}\,\text{eV}\)). In the case of nucleus‑induced EAS, the best‑fitting model estimation for the contribution of the PN\(\mu\) to the total ground‑level muon content is on the order of a few per cent, which agrees well with previous work~\cite{Muller:2018}.

The previously discussed muon puzzle refers to an observed excess of muons on the order of \(50\%\), compared to standard MC expectations~\cite{Velazquez:2023}. Assuming a linear scaling of the muon excess \(\Delta N_{\mu}\) with the cross section enhancement \(\Delta \bar{\sigma}_{\gamma A}\), one can use the characteristic ground-level ratio from the EGS4 model, \(N_{\muphNucl}^{\text{EGS4}}/N_{\mu} \sim 5\%\), to roughly estimate the enhancement required to produce this effect:
\begin{equation}
    \begin{aligned}
    &\frac{\Delta \bar{\sigma}_{\gamma A}(E_{\gamma A})}{\bar{\sigma}_{\gamma A}(E_{\gamma A} \mid \text{EGS4})} \sim \\
    &\sim 10 \times \left( \frac{\Delta N_{\mu} / N_{\mu}}{50\%} \right) \times \left( \frac{N_{\muphNucl}^{\text{EGS4}} / N_{\mu}}{5\%} \right)^{-1}.
    \end{aligned}
\end{equation}

\section{Conclusions}
\label{sec:concl}
This study develops and validates a heuristic model that establishes a linear mapping between high-energy PNR cross section and the muon content of extensive air showers. The model demonstrates applicability across a broad range of detection depths, primary particle types and energies, as well as various PNR models, with a characteristic absolute percentage error on the order of \(10\%\) at sufficiently large numbers of muons. The heuristic approach combines the universality of computationally intensive MC simulations with the computational efficiency of more limited analytical methods, providing a useful tool for quantitative analysis of PNRs in EAS physics.

The model can be applied to quantify how uncertainty in the high-energy PNR cross section impacts the muon puzzle. As demonstrated in this work, a one-order-of-magnitude enhancement of this cross section could substantially reduce its significance. Furthermore, the model can assess the impact of this uncertainty on the ability to discriminate between photon- and nucleus-induced EAS, a task which is likely to become obstructed at energies above \(10^{18}\,\text{eV}\). Additionally, the model provides a means to constrain the high-energy PNR cross section using EAS experimental data. These applications will be addressed in forthcoming publications.
\vspace{-\baselineskip}

\section*{Acknowledgments}
The author is indebted to L.\,Z. Dzhilavyan, D.\,V. Fedorova, N.\,N. Kalmykov, R.\,D. Monkhoev, A.\,A. Petrukhin, A.\,L. Polonski, G.\,I. Rubtsov, S.\,V. Troitsky, and the anonymous referee for valuable discussions and helpful suggestions. 

The author also expresses gratitude to the developers of the Python libraries Matplotlib~\cite{Matplotlib}, NumPy~\cite{NumPy}, Pathos~\cite{Pathos}, SciencePlots~\cite{SciencePlots}, Scikit-learn~\cite{scikit-learn}, and SciPy~\cite{SciPy} for providing their code as free and open-source software. 

The author thanks Theoretical Physics and Mathematics Advancement Foundation
“BASIS” for the student fellowships under the contract
24-2-10-39-1.

This work was supported by the Russian Science Foundation, grant 25-12-00333.

\section*{Data Availability}
The preprocessed MC simulation data supporting the findings of this study are openly available in the GitHub repository~\cite{Martynenko:2025}.

\appendix
\section{Hadronic cascade development beyond the Leading interactions tree}
\label{appendix:fudge-factor}

Recall that in this work, the transformation from the Leading interaction tree contribution \(G_{\leading}\) to the total \(G\) is encoded by a constant empirical correction factor \(\kappa_G = 5\).

To analyze the contribution of non-leading interactions to the resultant function \(G(E_{\gamma A},\,E_{\CR},\,X_{\mu})\), an oversimplified toy model of EAS development is introduced. The following assumptions are made: (i) \(X_{\mu}\) is sufficiently large such that \(\partial G_{\leading}/\partial X_{\mu}\) is negligible;  
(ii) in each leading interaction, the leading particle carries away exactly half of the interaction energy;  
(iii) the remaining interaction energy is equally distributed among non-leading particles;  
(iv) each charged non-leading particle produces its own Leading interactions tree and non-leading interactions, whose total contribution can be approximated by \(\kappa_G \times G_{\leading}\). Under these assumptions and using Eq.~\eqref{eq:fudge-factor}, the following relation can be derived:
\begin{equation}
\begin{aligned}
&\kappa_G \times G_{\leading}(E_{\gamma A},\,E_{\CR},\,X_{\mu}) = G_{\leading}(E_{\gamma A},\,E_{\CR},\,X_{\mu}) + \\
&+ \sum_{k=1}^{k_{\max}} \Bigg[\mathcal{M}_{\text{ch}}\left(2^{-k}E_{\CR}\right) \times \kappa_G \times \\
&\qquad\times G_{\leading}\left(E_{\gamma A},\,\frac{2^{-k}E_{\CR}}{\mathcal{M}_{\text{tot}}\left(2^{-k}E_{\CR}\right)},\,X_{\mu}\right)\Bigg],
\end{aligned}
\end{equation}
where \(\mathcal{M}_{\text{ch}}\) and \(\mathcal{M}_{\text{tot}}\) are, respectively, the energy-dependent charged and total multiplicity.
Additional assumptions are introduced to simplify the analysis: (v) \(G_{\leading}\) is approximately proportional to the primary cosmic-ray proton energy, i.e., \(G_{\leading}/E_{\CR} \to \operatorname{const}\);  
(vi) the charged-to-total multiplicity ratio \(0 < \mathcal{M}_{\text{ch}}/\mathcal{M}_{\text{tot}} < 1\) is nearly constant, with \(\mathcal{M}_{\text{tot}} \gg 1\) for all relevant interaction energies. Under assumptions (v) and (vi), Eq.~\eqref{eq:fudge-factor-toy-model} becomes:
\begin{equation}
\label{eq:fudge-factor-toy-model}
\kappa_G = 1 + \kappa_{G} \times \frac{\mathcal{M}_{\text{ch}}}{\mathcal{M}_{\text{tot}}} \sum_{k=1}^{k_{\max}}\frac{1}{2^k}.
\end{equation}
For brevity, define:
\begin{equation}
S_G = \frac{\mathcal{M}_{\text{ch}}}{\mathcal{M}_{\text{tot}}} \sum_{k=1}^{k_{\max}}\frac{1}{2^k} < 1,
\end{equation}
and solve Eq.~\eqref{eq:fudge-factor-toy-model} for \(\kappa_G\):
\begin{equation}
\kappa_G = \frac{1}{1 - S_G}.
\end{equation}

The empirical correction factor \(\kappa_G\) exhibits a strong dependence on \(S_G\). Specifically, the partial derivative \(\partial \kappa_G / \partial S_G\) evaluates to approximately \(25\) at the reference value \(\kappa_G = 5\) adopted in this study. This high sensitivity underscores the critical role of \(S_G\) in determining \(\kappa_G\).

However, the exact value of \(S_G\) cannot be determined without a detailed model of hadronic interactions and relaxation of most of the underlying assumptions.

Notably, the Heitler–Matthews model~\cite{Matthews:2005} predicts \(S_G = 2/3\), which yields \(\kappa_G = 3\). This prediction agrees with simulation results within tens of percent, indicating qualitative consistency between the simplified model and numerical experiments.

\bibliography{photonucl}
\end{document}